\documentclass[conference,compsoc]{IEEEtran}

\ifCLASSOPTIONcompsoc
  \usepackage[nocompress]{cite}
\else
  \usepackage{cite}
\fi

\usepackage{tabularx}
\usepackage{longtable}
\usepackage{booktabs} 
\usepackage{threeparttable} 

\usepackage[utf8]{inputenc}
\usepackage{array}          
\usepackage{graphicx}       
\usepackage{pifont}         
\usepackage{booktabs}       
\usepackage{multirow}       
\usepackage[table]{xcolor}  
\usepackage{adjustbox}      

\usepackage{amsmath}
\usepackage{wasysym}

\usepackage{url}
\usepackage[hidelinks]{hyperref}
\usepackage{xurl}
\usepackage{cleveref}

\usepackage[inline]{enumitem} 

\usepackage[acronym]{glossaries} 
\glsdisablehyper

\usepackage{rotating}
\usepackage{makecell}
\usepackage{xparse}
\usepackage{nicematrix}
\usepackage{tikz}
\usetikzlibrary{calc}
\usepackage[colorinlistoftodos]{todonotes}

\usepackage{harveyballs}

\usepackage{etoolbox}

\usepackage{microtype}
\microtypecontext{spacing=nonfrench}

\usepackage{tikz}
\usepackage[framemethod=TikZ]{mdframed}
\usepackage{tcolorbox}

\NewEnviron{summarybox}[1]{%
  \begin{mdframed}[%
    skipabove=14pt plus 3pt minus 2pt,
    skipbelow=3pt plus 3pt,
    middlelinecolor=black,
    middlelinewidth=1pt,
    frametitle={\smash{\tcbox[
      on line, tcbox raise=0pt, arc=0mm,
      boxsep=0pt, left=8pt, right=8pt, top=2pt, bottom=2pt,
      colback=black, colframe=black, boxrule=1pt]%
      {\small\color{white}#1}}}, 
    frametitleaboveskip=5pt,
    frametitlebelowskip=0pt,
    innertopmargin=5pt,
    frametitlerule=false,
    frametitlerulewidth=0pt,
    frametitlebackgroundcolor=none]
    \BODY 
  \end{mdframed}
}

\newtcolorbox{designimplication}{%
  colback=blue!5, colframe=blue!40,
  boxrule=0.5pt, arc=0.6mm,
  left=6pt, right=6pt, top=4pt, bottom=4pt,
  before skip=6pt plus 2pt minus 1pt, after skip=6pt plus 2pt minus 1pt,
  fontupper=\small,
  before upper={\textbf{Design Implication:}\space}%
}
\newtcolorbox{researchopportunity}{%
  colback=orange!8, colframe=orange!55,
  boxrule=0.5pt, arc=0.6mm,
  left=6pt, right=6pt, top=4pt, bottom=4pt,
  before skip=6pt plus 2pt minus 1pt, after skip=6pt plus 2pt minus 1pt,
  fontupper=\small,
  before upper={\textbf{Research Opportunity:}\space}%
}

\usepackage{circledsteps}

\newcommand{\eolquote}[2]{\textit{``#1''} (#2)}

\newcommand{\squote}[1]{\textit{``#1''}}

\ExplSyntaxOn
\NewExpandableDocumentCommand { \ValuePlusOne } { m } 
  { \int_eval:n { \int_use:c { c @ #1 } + 1 } }
\NewExpandableDocumentCommand { \Sec } { m } 
  { \fp_eval:n { secd ( #1 ) } }
\NewDocumentCommand { \Rot } { m }
  { 
    \hbox_to_wd:nn { 1 em }
      { 
        \hbox_overlap_right:n 
          { 
            \skip_horizontal:n { \fp_to_dim:n { 7 * cosd (\Angle) } } 
            \rotatebox{\Angle}{#1}
          } 
      } 
  }
\ExplSyntaxOff

\NewDocumentCommand { \MixedRuleShift } { m }
  {
    \begin{tikzpicture}
    \coordinate (a) at (2-|#1);
    \coordinate (aa) at ($(a)-(.15,0)$);
    \coordinate (b) at (1-|#1) ;
    \coordinate (bb) at ($(b)-(.15,0)$);
    \draw (aa) -- ($(aa)!\Sec{90-\Angle}!\Angle-90:(bb)$) ;
    \draw (aa) -- (\ValuePlusOne{iRow}-|aa) ;
    \end{tikzpicture}
  }

\newcommand{\usecase}[1]{\Circled[fill color=black, inner color=white, inner ysep=4pt, inner xsep=4pt]{\footnotesize \textnormal{#1}}}

\newacronym{AI}{AI}{Artificial Intelligence}
\newacronym{CI/CD}{CI/CD}{Continuous Integration / Continuous Deployment}
\newacronym{CISO}{CISO}{Chief Information Security Officer}
\newacronym{DGA}{DAG}{Domain Generation Algorithms}
\newacronym{GDPR}{GDPR}{General Data Protection Regulation}
\newacronym{DORA}{DORA}{Digital Operational Resilience Act}
\newacronym{GenAI}{GenAI}{Generative Artificial Intelligence}
\newacronym{UEBA}{UEBA}{User Entity Behavior Analytics}
\newacronym{HITL}{HITL}{Human-in-the-Loop}
\newacronym{HMI}{HMI}{Human-Machine-Interface}
\newacronym{IR}{IR}{Incident Response}
\newacronym{IRB}{IRB}{Institutional Review Board}
\newacronym{IRR}{IRR}{Inter-Rater Reliability}
\newacronym{LLM}{LLM}{Large Language Model}
\newacronym{ML}{ML}{Machine Learning}
\newacronym{PII}{PII}{Personal Identifiable Infomation}
\newacronym{RAG}{RAG}{Retrieval-Augmented Generation}
\newacronym{ROI}{ROI}{Return of Invest}
\newacronym{RQ}{RQ}{Research Question}
\newacronym{SOC}{SOC}{Security Operations Center}
\newacronym{MSOC}{MSOC}{managed SOC}
\newacronym{SOP}{SOP}{Standard Operating Procedures}
\newacronym{TI}{TI}{Threat Intelligence}
\newacronym{TTP}{TTP}{Tactics, Techniques, and Procedures}
\newacronym{LLMs}{LLMs}{Large Language Models}
\newacronym{EDR}{EDR}{Endpoint Detection and Response}
\newacronym{SIEM}{SIEM}{Security Information and Event Management}
\newacronym{SOAR}{SOAR}{Security Orchestration, Automation and Response}
\newacronym{XAI}{XAI}{Explainable AI}

\begin{document}


\title{From Chasing Ghosts to Missed Attacks: Perspectives and Perceptions of SOC Practitioners on LLM Integration, Risks, and Readiness}

\author{\IEEEauthorblockN{
Jonas Thurner\IEEEauthorrefmark{1},
Nadine Jost\IEEEauthorrefmark{2},
Stefan Albert Horstmann\IEEEauthorrefmark{3},\\
Fabian Ising\IEEEauthorrefmark{4},
Lea Gr\"ober\IEEEauthorrefmark{5},
Alena Naiakshina\IEEEauthorrefmark{3},
Sebastian Schinzel\IEEEauthorrefmark{1}\IEEEauthorrefmark{4}}
\IEEEauthorblockA{\IEEEauthorrefmark{1}FH M\"unster\\
j.thurner@fh-muenster.de, schinzel@fh-muenster.de}
\IEEEauthorblockA{\IEEEauthorrefmark{2}Ruhr University Bochum\\
nadine.jost@rub.de}
\IEEEauthorblockA{\IEEEauthorrefmark{3}University of Cologne\\
stefan.horstmann@uni-koeln.de, alena.naiakshina@uni-koeln.de}
\IEEEauthorblockA{\IEEEauthorrefmark{4}Fraunhofer SIT and National Research Center for Applied Cybersecurity ATHENE\\
fabian.ising@sit.fraunhofer.de}
\IEEEauthorblockA{\IEEEauthorrefmark{5}ICSI, UC Berkeley\\
lgrober@icsi.berkeley.edu}}

\maketitle


\begin{abstract}    
    \glspl{SOC} process large volumes of security events, requiring analysts to accurately detect and assess ongoing cyberattacks under time pressure. Recent advances in \glspl{LLM} suggest potential benefits for security operations, yet their practical suitability for real-world SOC workflows remains poorly understood. To address this gap, we conducted 25 semi-structured interviews with SOC practitioners who had prior experience with LLMs, complemented by interactive scenarios to anticipate challenges and identify opportunities for the responsible integration of LLM-based tools into SOC workflows.
    We identified 15 LLM use cases grouped into six functional categories. While \glspl{LLM} are valued for automating repetitive, low-level tasks such as report automation, practitioners rate high-impact tasks such as incident analysis as not yet feasible, reporting limitations in technical depth, context awareness, and organization-specific knowledge. They attribute these limitations less to the models than to the readiness of their SOCs and to human factors driving over-reliance.
    Despite concerns, practitioners express a strong willingness to adopt \glspl{LLM}, describing competitive pressure that leaves few alternatives. This work contributes an empirical, practitioner-driven analysis of \gls{LLM} use across SOC roles and organizations and derives concrete design and integration requirements for human-centered, operationally safe LLM-assisted security operations.
\end{abstract}

\glsresetall
\section{Introduction}

\glspl{SOC}, operated either in-house or as managed services (MSOCs), are organizational units run by medium-to-large enterprises, governmental organizations, and critical infrastructure providers to monitor, detect, and respond to cyber threats~\cite{WhatIsaS89:online}.
However, the effectiveness of SOCs heavily relies on the human expertise of the professionals who operate them~\cite{vielberth_security_2020,SecurityOperationsCentera}.
SOC practitioners span a broad range of roles within the SOC.
At their core, security analysts monitor networks and endpoints, analyze alerts, and identify threats. Security engineers design and maintain the technical infrastructure, deploying intrusion detection systems and automated response mechanisms.
Red team specialists contribute to stronger defenses by identifying weaknesses and blind spots in detection logic.
Overseeing these efforts, the SOC manager or \gls{CISO} aligns operational activities with organizational risk profiles and regulatory compliance requirements. 

However, SOC analysts frequently experience cognitive overload and alert fatigue due to an overwhelming influx of alerts, many of which are false positives~\cite{nepal_burnout_2024,sundaramurthy_human_2015,sundaramurthy_tale_2014}. Consequently, analysts spend substantial time manually verifying these false positives, increasing the risk of overlooking genuine threats~\cite{alahmadi99FalsePositives2022f,sundaramurthy_human_2015,sundaramurthy_tale_2014}. Simultaneously, security engineers are often distracted from strategic tasks by repetitive, low-level duties such as fine-tuning detection rules to minimize false positives~\cite{nepal_burnout_2024,alahmadi99FalsePositives2022f}. Ultimately, these inefficiencies strain SOC resources, impairing their ability to proactively respond to evolving threats and underscoring the critical need for innovative solutions to enhance SOC effectiveness.

While \gls{AI} systems are already deployed in SOCs to detect and classify anomalies~\cite{mink_everybodys_2023,oesch_assessment_2020}, the recent rise of \glspl{LLM} has introduced new possibilities to support practitioners and automate security workflows~\cite{freitas_ai-driven_2024,ferrag_revolutionizing_2024,ferrag_generative_2025,saddiExamineRoleGenerative2024a,tseng_using_2024,mezziLargeLanguageModels2025,elsharef_facilitating_2024,hartsock_towards_2024,hassanin_comprehensive_2024,pasupuleti_cyber_2023,xu_large_2024,AISecurityTrends,GenerativeAIWhat}. However, deploying LLMs in security-specific contexts introduces new risks. Prior research has shown that \glspl{LLM} may overlook critical security details or generate plausible yet incorrect security recommendations~\cite{mezziLargeLanguageModels2025,hassanin_comprehensive_2024,pasupuleti_cyber_2023}.
While LLM integration in SOCs is still emerging, the associated opportunities and risks call for guidance on responsible adoption.
To this end, we conducted a qualitative study with 25 SOC practitioners spanning the full spectrum of SOC roles~\cite{vielberth_security_2020}. To support participants in articulating abstract concepts and encourage critical reflection, we combined semi-structured interviews with interactive brainstorming and visualization tasks~\cite{GraphicElicitationUsing,CreativityContextUpdate}. We addressed the following \glspl{RQ}:

\begin{itemize}[align=left, widest={RQ1:},leftmargin=*]
    \item[\textbf{RQ1:}] Which \textit{use cases} do practitioners identify for \glspl{LLM} in SOCs?

    \item[\textbf{RQ2:}] What are the \textit{perceived technical and operational challenges} of integrating \glspl{LLM} in SOC workflows? 

    \item[\textbf{RQ3:}] How do practitioners perceive \textit{LLM-specific risks} in the context of security operations and what \textit{countermeasures} do they propose?

\end{itemize}

Our study surfaces 15 use cases across six categories and reveals a gap between the envisioned capabilities of LLMs and the operational realities of SOCs: most practitioners value LLMs for repetitive, low-level tasks such as reporting and \gls{TI} summarization, but a majority reject them for core security decisions such as detection and incident analysis. They described a shift in concern from the false positives that drive conventional SOC work~\cite{nepal_burnout_2024, alahmadi99FalsePositives2022f} toward the false negatives produced by over-reliance, as they delegate more to LLMs.
Overall, our study contributes a practitioner-driven, empirical characterization of LLM integration across multiple SOCs and roles, highlighting over-reliance as a central concern and providing practitioner-grounded insights into the design of  assistive LLM systems for security operations.

\section{Related Work}

In this section, we present prior research on human factors, \gls{ML}, and Usable Security of ML in SOCs. 

\subsection{Human Factors in SOCs}
Related work on human factors in SOCs reveals a range of challenges affecting practitioner performance and well-being, including alert fatigue, burnout, cognitive overload, and poor workforce management~\cite{alahmadi99FalsePositives2022f, sundaramurthy_human_2015, kokulu_matched_2019, nepal_burnout_2024, schlette_comparative_2021, sundaramurthy_tale_2014, sundaramurthy_turning_2016, tariqAlertFatigueSecurity2025, jack_tilbury_humans_2024, vielberth_security_2020, yang_true_2024}.
Burnout in particular is driven by workload, insufficient management support, and the volume of false-positive alerts~\cite{nepal_burnout_2024, alahmadi99FalsePositives2022f}.
Unlike most studies, Kokulu et al.~\cite{kokulu_matched_2019} found that practitioners did not view false positives as a primary issue, instead identifying organizational misalignments, such as management's focus on metrics versus analysts' emphasis on usability and operational constraints.
Other work points to human-centered and socio-technical design as key to stronger defenses~\cite{helkala_supporting_2018,gerontakis_security_2023}, to organizational support and tool integration for mitigating SOC skill shortages~\cite{nyre-yu_identifying_2021}, to outsourcing models for human-resource limitations~\cite{shah_outsourcing_2020}, and to the impact of \gls{IR} playbooks on analyst performance~\cite{schlette_comparative_2021,schlette_you_2024}.
Stevens et al.~\cite{stevens_how_2022} found many such playbooks lacked clarity and detail, especially for less experienced analysts, urging iterative refinement. We extend this line of research to the LLM era, examining how practitioners perceive LLM capabilities, limitations, risks, and anticipated changes in real-world integration.

\subsection{Machine Learning in SOCs}

ML, and \glspl{LLM} in particular, are a promising but controversial technology. Recent studies investigate LLMs for SOC-specific tasks such as \gls{TI} processing~\cite{saddiExamineRoleGenerative2024a,tseng_using_2024,mezziLargeLanguageModels2025}, intrusion detection~\cite{ferrag_revolutionizing_2024,zhang_caravan_2024}, threat modeling~\cite{elsharef_facilitating_2024}, investigation~\cite{freitas_ai-driven_2024,hartsock_towards_2024}, and triage~\cite{freitas_ai-driven_2024}, alongside systematic reviews of LLM applications in SOCs~\cite{ferrag_generative_2025,hasanov_application_2024,hassanin_comprehensive_2024,pasupuleti_cyber_2023,xu_large_2024}. While largely encouraging, several studies flag risks: Mezzi et al.~\cite{mezziLargeLanguageModels2025} find that LLMs omit critical details, making them too risky for \glspl{SOC}, where a small oversight can let adversaries go undetected, while Hassanin et al.~\cite{hassanin_comprehensive_2024} and Pasupuleti et al.~\cite{pasupuleti_cyber_2023} warn about hallucinations. We complement these technical studies with a user-centered approach, exploring practitioner perspectives and the constraints of real-world security workflows.

\subsection{Usable Security of ML in SOCs}

Practitioner studies report low adoption of AI/ML in SOCs and attribute it to unreliable alarms and models that cannot be adequately tested for high-security environments~\cite{vermeerAlertAlchemySOC2023, alahmadi99FalsePositives2022f}. \glspl{LLM} differ from these tools in that they are readily accessible, and it is tempting to delegate security questions to them. Klemmer et al.~\cite{klemmerUsingAIAssistants2024} found that software professionals already use AI assistants for security-related development despite quality and security concerns, which suggests \glspl{LLM} may see similar uptake in SOCs.

Earlier practitioner studies of SOC tooling are pre-LLM, analyst-centric, and detection-focused~\cite{oesch_assessment_2020, mink_everybodys_2023}. Hahn et al.~\cite{NonDisruptiveDisruptionEmpiricalb} co-developed a companion platform in one SOC. By contrast, we examine various \gls{LLM} use cases and associated risks, focusing on general-purpose models because security-specific tools are typically built on top of them and may inherit their capabilities and limitations. Singh et al.~\cite{singhLLMsSOCEmpirical2025a} found that 31\% of MSSP analyst queries involve command understanding. Our findings link this use case to practitioner concerns and potential risks (\Cref{tab:related_work}). We study \glspl{LLM} across 17 organizations, the full SOC role spectrum~\cite{vielberth_security_2020}, and the different use cases, pairing practitioner perceptions with a structured assessment of \gls{LLM}-specific risks.

\begin{table}[t]
    \caption{Empirical practitioner studies of AI/ML and \gls{LLM} tooling in SOCs, compared across generative-AI focus, breadth of LLM use cases, SOC and practitioner coverage, and role diversity.}
    \label{tab:related_work}
    \centering
    \footnotesize
    \setlength{\tabcolsep}{4.5pt}
    \resizebox{\columnwidth}{!}{%
    \begin{tabular}{l c c c c c}
        \toprule
        Study & GenAI & LLM use cases & Scope & Sample size & Roles \\
        \midrule
        Oesch et al.~\cite{oesch_assessment_2020}            & no  & none     & single-org & 6    & analysts \\
        Alahmadi et al.~\cite{alahmadi99FalsePositives2022f} & no  & none     & cross-org  & 21   & all \\
        Mink et al.~\cite{mink_everybodys_2023}              & no  & none     & cross-org  & 18   & all \\
        Vermeer et al.~\cite{vermeerAlertAlchemySOC2023}     & no  & none     & cross-org  & 17   & all \\
        Hahn et al.~\cite{NonDisruptiveDisruptionEmpiricalb} & yes & companion & single-org & co-dev & all \\
        Singh et al.~\cite{singhLLMsSOCEmpirical2025a}       & yes & multiple & single-org & none & analysts \\
        \midrule
        \rowcolor[gray]{0.9}
        \textbf{This work} & \textbf{yes} & \textbf{multiple} & \textbf{cross-org} & \textbf{25} & \textbf{all} \\
        \bottomrule
    \end{tabular}%
    }
\end{table}

\section{Methodology}

We conducted 25 semi-structured interviews with SOC practitioners who were screened with a demographic survey to ensure prior experience with LLMs.
At the time of the study, it was unclear whether and how LLMs were being used in SOCs. We therefore chose an open and exploratory interview design, which allowed us to capture a wide range of insights and perspectives.
We therefore enhanced the interviews by interactive Miro board~\cite{Miro} sections incorporating brainstorming and flowchart creation for in-depth, participant-driven exploration of novel topics as they emerged.

\subsection{Study Design}

We developed an interview guide grounded in our RQs, which was refined through discussions with researchers experienced in security and human factors to ensure comprehensive coverage and clear phrasing. To test the interview guide, we conducted three pre-tests with security researchers taking on the persona of a SOC analyst, checking flow, question order, and the Miro setup. The following adjustments were made to the interview guide: 
(1) We shifted the interview guide from a research question-based guideline to a topic-focused structure for smoother flow and more cohesive responses.
(2) To capture the participants' thoughts in detail, we replaced general questions (e.g., \squote{What are the challenges of integrating \glspl{LLM}?}) with more specific questions (e.g., \squote{What are the technical challenges of \gls{LLM} integrations?}). Interviews were conducted between January 2025 and April 2026, offering participants the choice of English or German. All interviews were conducted by the first author via Zoom~\cite{VideoConferencingWeb}, advertised as 90-minute sessions, averaging 79 minutes, with a 5-minute break to counteract fatigue effects. We collected a total of 32 hours and 56 minutes of interview material. 

\subsection{Interview Guideline}
\label{met:interviewguide}

\begin{figure}[t]
  \centering
  \includegraphics[width=0.8\linewidth]{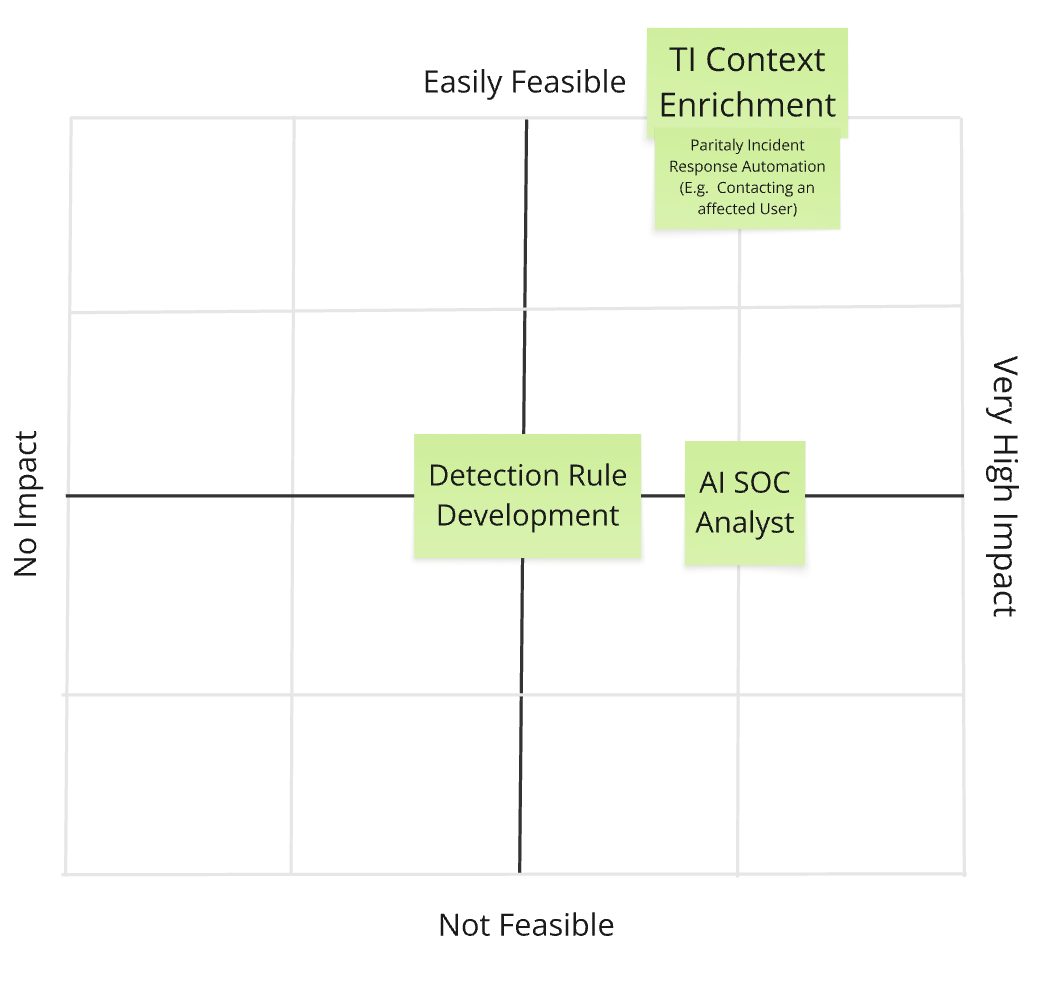}
  \caption{Example of the impact and feasibility assessment.}
  \label{fig:ImpactFeasibilityExample}
\end{figure}

\begin{figure}[t]
  \centering
  \includegraphics[width=0.95\linewidth]{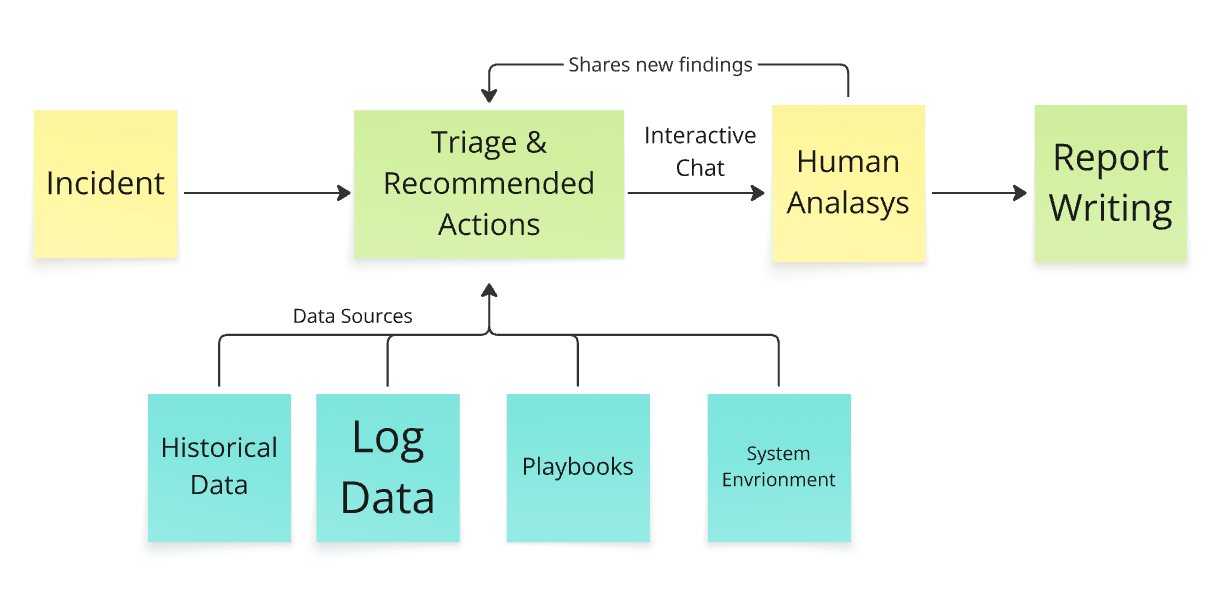}
  \caption{Example flowchart created by a participant. Green boxes indicate LLM-assisted tasks.}
  \label{fig:MiroBoardLLMSOCIntegrationExample}
\end{figure}


Before each interview, we obtained written consent and conducted a short demographic survey (see Section~\ref{methods:participants}). The interview guide, recruitment material, and our demographic survey are available online (cf.~\Cref{sec:availability}). To establish a shared understanding, we introduced the term \textit{general-purpose LLM} as a reference point for models such as ChatGPT, Gemini, and Llama. While the discussion was not limited to these systems, this distinction helped participants differentiate them from security-specific or embedded tools such as Microsoft Security Copilot. Afterwards, we obtained oral consent and started the recording to discuss the following topics.

\textit{General Questions:} We asked participants about their daily work routines and challenges, their familiarity with \glspl{LLM}, including perceived strengths and weaknesses. We also explored their predictions for \gls{LLM} development in SOCs. Additionally, we inquired about their current use of \glspl{LLM} in security operations, the challenges they have encountered and how they addressed them, the areas where \glspl{LLM} have succeeded or failed, and the reasons for these outcomes. This section served both as an ice-breaker prior to the interactive tasks, and as a means to verify participants' statements in the demographic screening survey.

\textit{Brainstorming (RQ1):} Next, we conducted a brainstorming session on a Miro board~\cite{Miro}, encouraging participants to share potential LLM applications in security operations without any limitations. Some reflected on their own practical experiences, while others proposed novel applications.
Each participant worked with an individual Miro board to ensure responses were not influenced by others. We asked them to think aloud as they added their ideas to the board. We provided optional questions to stimulate brainstorming if needed. 
This open-ended elicitation allowed RQ1 to remain participant-driven, rather than constraining responses to researcher-defined use cases.

\textit{Impact and Feasibility Assessment (RQ1):} Subsequently, we asked the participant to place their identified use cases on a grid. We asked them to think aloud and explain (1) their perceived impact of the solution and (2) its feasibility (see~\Cref{fig:ImpactFeasibilityExample} for an example). During this step, we also noted, for each application, whether the participant grounded it in hands-on experience or envisioned the use case.
This assessment helped differentiate promising use cases from those perceived as difficult to realize in practice.

\begin{figure}[tb]
  \centering
  \includegraphics[width=0.95\linewidth]{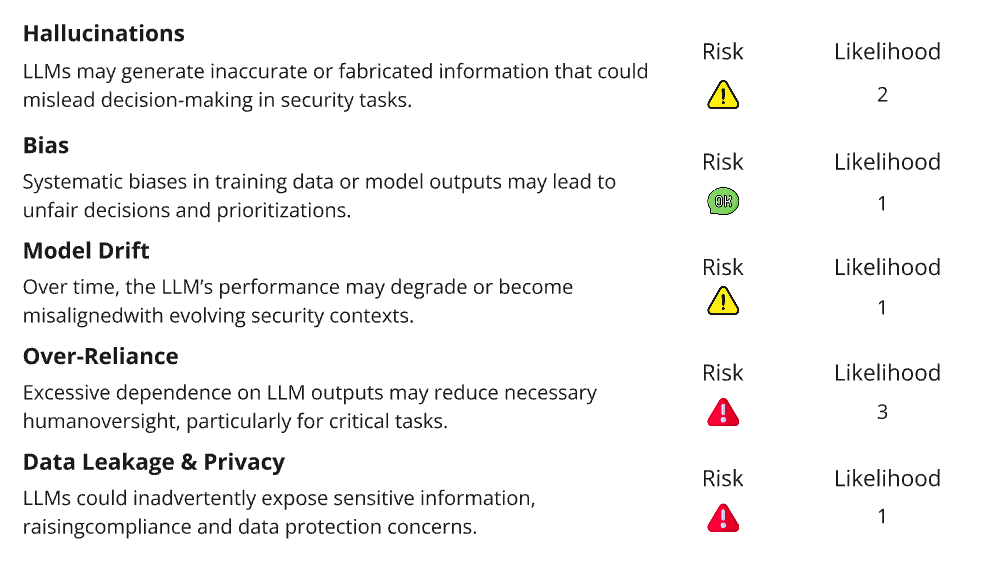}
  \caption{Illustrative example of the risk assessment. Risk was defined as the consequence if the risk materializes and is represented by icons: green (1 = low), yellow (2 = medium), and red (3 = high).}
  \label{fig:MiroBoardRiskAssessmentExample}
\end{figure}

\textit{Integration Flowchart (RQ2):} 
After that, we prompted participants to create a flowchart visualizing their ideal \gls{LLM}-driven \gls{SOC} workflow (see~\Cref{fig:MiroBoardLLMSOCIntegrationExample} for an example). 
This exercise enabled an in-depth exploration of one use case that each participant deemed impactful based on their own judgment, balancing exploratory depth with the available time.
Participants were asked to think aloud. If necessary, participants were provided technical assistance, e.g., when drawing arrows. The first author used a checklist~(see \Cref{sec:availability}) that the research team had drafted as probe questions, informed by BPMN\footnote{Business Process Model and Notation, a standard graphical notation for specifying business processes~\cite{allweyerBPMN20Business2020}.}. 
The first author used the checklist to support participants in covering key aspects. For example, when a participant described a process step involving an LLM, the interviewer could ask what input the LLM received and what output it was expected to produce. These questions helped ensure that relevant process details were captured.

\textit{Critical Engagement (RQ2):} Next, we discussed the flowchart, focusing on technical, organizational, and regulatory challenges. We questioned the viability of the proposed solution. Participants were asked to challenge their own design by arguing against it to identify essential pain points and where human involvement remained critical. This encouraged deeper reflection on the proposed solution and helped counteract overly positive assessments.

\textit{\gls{LLM}-Specific Risk Assessment (RQ3):} After participants had discussed the challenges they identified themselves, we presented them with five \gls{LLM}-specific risks (\Cref{tab:riskDefinitions}). Participants were security experts but not necessarily \gls{ML} experts, so the point of this stage was to introduce \gls{LLM}-specific risks they might not have in mind.
We asked them to: (1) assess each risk’s impact and likelihood while thinking aloud and (2) suggest ways to detect or mitigate it within the context of the designed flowchart (see~\Cref{fig:MiroBoardRiskAssessmentExample}).
We selected these five risks in discussion with two senior AI/ML researchers. The risks were chosen to fit within the scope of the 90-minute interviews and do not represent an exhaustive set. We selected risks that represent potential consequences of integrating \glspl{LLM} into SOC workflows, rather than specific attack techniques against the LLM. We therefore excluded attack mechanisms such as prompt injection, or data poisoning. However, participants raised them continuously unprompted. 
We therefore report this as an additional emerging risk in \Cref{res:rq3:riskPerceptions}. 
Afterwards, we invited participants to name any further risks.
\begin{table}[t]
    \caption{Definitions of the five predefined \gls{LLM}-specific risks as presented to participants.}
    \label{tab:riskDefinitions}
    \begin{threeparttable}
    \centering
    \setlength{\tabcolsep}{5pt}
    \scriptsize
    \begin{NiceTabular}{@{}p{1.35cm} X[l] c@{}}[width=\linewidth]
      \toprule
      \textbf{Risk} & \textbf{Definition presented to participants} & \textbf{Source} \\
      \midrule
      \rowcolor[gray]{0.9}
      Hallucinations &
      LLMs may generate inaccurate or fabricated information that could mislead decision-making in security tasks. &
      \cite{huangSurveyHallucinationLarge2025,WhatAreAI2024} \\

      Bias &
      Systematic biases in training data or model outputs may lead to unfair decisions and prioritizations. &
      \cite{echterhoffCognitiveBiasDecisionMaking2024,DefiningEvaluatingPolitical2026} \\

      \rowcolor[gray]{0.9}
      Model Drift &
      Over time, the LLM's performance may degrade or become misaligned with evolving security contexts. &
      \cite{abdelnabiGetMyDrift2025,WhatModelDrift2024} \\

      Over-Reliance &
      Excessive dependence on LLM outputs may reduce necessary human oversight, particularly for critical tasks. &
      \cite{spathariotiEffectsLLMbasedSearch2025,mvorvoreanuOverrelianceAIRisk} \\

      \rowcolor[gray]{0.9}
      Data Leakage \& Privacy &
      LLMs could inadvertently expose sensitive information, raising compliance and data protection concerns. &
      \cite{zhouLessLeakBenchFirstInvestigation2025,WhatDataLeak} \\
      \bottomrule
    \end{NiceTabular}
    \end{threeparttable}
\end{table}

In the end, we invited participants to share final thoughts, thanked them for their time, and concluded the session.

\subsection{Participants}
\label{methods:participants}

\begin{table*}[ht!]

\scriptsize
\centering
\def\Angle{45}

\newcommand{\rotheader}[2]{\def\Angle{#1}\Rot{#2}}
\caption{Overview of participants' demographics.}
\label{tab:overview}

\begin{threeparttable}
\centering
\setlength{\tabcolsep}{5pt}
\begin{NiceTabular}{lcllcrr lccc w{c}{1.2cm}ccc w{l}{.35cm} w{c}{2.2cm}}
\CodeBefore
\rowcolors{3}{gray!25}{white}
\Body

\toprule
\rotheader{45}{\textbf{ID}}& 
\rotheader{45}{\textbf{Location}}& 
\rotheader{45}{\textbf{Education}}&
\rotheader{45}{\textbf{Occupation}}&
\rotheader{45}{\textbf{SOC type}}&
\multicolumn{1}{c}{\rotheader{45}{{\textbf{Experience} (yrs)}}}&
\multicolumn{1}{c}{\rotheader{45}{{\textbf{Environment} (ppl)}}}&
 \rotheader{45}{\textbf{LLM Deployment}} &
 \rotheader{45}{Commercial Cloud\textsuperscript{1}} &
 \rotheader{45}{Private Cloud\textsuperscript{2}} &
 \rotheader{45}{Self-Hosted\textsuperscript{3}} &
 \rotheader{45}{\textbf{LLM Experience}} &
 \rotheader{45}{Interactive Chat} &
 \rotheader{45}{Scripting} &
 \rotheader{45}{Embedded} & 
 \rotheader{45}{Autonomous Agent} &
 \multicolumn{1}{l}{\rotheader{45}{\textbf{Usage Frequency}}}
  \\
\midrule
A1 & EU & M. Sc.  & SOC Analyst & In-house & + 20   & $<$ 1,000 & & \LEFTcircle & \Circle & \Circle & & \LEFTcircle & \Circle & \Circle & \Circle & monthly \\
A2 & EU & -  & SOC Analyst & In-house & 11-20   & $<$  10,000 & & \CIRCLE & \CIRCLE & \Circle & & \CIRCLE & \CIRCLE & \CIRCLE & \Circle & multiple/week \\
A3 & EU & M. Sc.  & SOC Analyst & In-house & 6-10   & $<$  10,000 & & \LEFTcircle & \Circle & \Circle & & \LEFTcircle & \Circle & \CIRCLE & \Circle & monthly \\
A4 & EU & M. Sc.  & SOC Analyst & Managed & 11-20   & $<$  1,000 & & \CIRCLE & \CIRCLE & \CIRCLE & & \CIRCLE & \CIRCLE & \LEFTcircle & \Circle & multiple/week \\
A5 & EU & M. Sc.  & SOC Analyst & In-house & 2-5 & $<$  1,000 & & \Circle & \CIRCLE & \Circle & & \CIRCLE & \Circle & \Circle & \Circle & weekly \\
A6 & US & B. Sc.  & SOC Analyst & In-house & 2-5 & $<$  10,000 & & \CIRCLE & \Circle & \Circle & & \CIRCLE & \CIRCLE & \Circle & \Circle & multiple/week \\
A7 & EU & M. Sc & SOC Analyst  & Managed & 2-5 & $<$  1,000 & & \Circle & \Circle & \LEFTcircle & & \Circle & \Circle & \LEFTcircle & \LEFTcircle & monthly \\
E1 & EU & B. Sc.  & Sec. Engineer & Managed & 2-5   & $<$ 200 & & \CIRCLE & \Circle & \CIRCLE & & \CIRCLE & \CIRCLE & \LEFTcircle & \LEFTcircle & daily \\
E2 & EU & Ph. D.  & Sec. Engineer & Managed & 2-5   & $<$ 200 &  & \CIRCLE & \Circle & \Circle & & \CIRCLE & \LEFTcircle & \LEFTcircle & \LEFTcircle & multiple/week \\
E3 & EU & M. Sc.  & Sec. Engineer & In-house & 0-2   & $<$ 1,000 & & \CIRCLE & \Circle & \Circle & & \CIRCLE & \Circle & \Circle & \Circle & multiple/week \\
E4 & EU & M. Sc.  & Sec. Engineer & In-house & 0-2   & $<$  50 & & \CIRCLE & \Circle & \CIRCLE & & \CIRCLE & \LEFTcircle & \Circle & \LEFTcircle & daily \\
E5 & EU & B. Sc.  & Sec. Engineer & In-house & 6-10   & $<$  10,000 & & \CIRCLE & \LEFTcircle & \Circle & & \CIRCLE & \CIRCLE & \Circle & \Circle & multiple/week \\
E6 & US & M. Sc.  & Sec. Engineer & Managed & 2-5   & $<$  10,000 & & \CIRCLE & \LEFTcircle & \CIRCLE & & \CIRCLE & \CIRCLE & \LEFTcircle & \CIRCLE & multiple/week \\
E7 & EU & B. Sc.  & Sec. Engineer & Managed & 11-20   & $>$  10,000 & & \CIRCLE & \CIRCLE & \CIRCLE & & \CIRCLE & \CIRCLE & \CIRCLE & \CIRCLE & daily \\
E8 & EU & M. Sc.  & Sec. Engineer & Managed & 6-10 & $>$  10,000 & & \LEFTcircle & \LEFTcircle & \Circle & & \LEFTcircle & \Circle & \CIRCLE & \Circle & multiple/week \\
E9 & EU & M. Sc & Sec. Engineer  & In-house & 6-10  & $<$  10,000  & & \CIRCLE & \Circle & \CIRCLE & & \CIRCLE & \CIRCLE & \Circle & \Circle & multiple/week \\
R1 & EU & Ph. D.  & Red Teamer & In-house & 11-20   & $<$ 200 & & \CIRCLE & \LEFTcircle & \LEFTcircle & & \CIRCLE & \LEFTcircle & \Circle & \Circle & daily \\
R2 & EU & Ph. D.  & Red Teamer & In-house & 6-10   & $<$ 1,000 & & \CIRCLE & \Circle & \LEFTcircle & & \CIRCLE & \Circle & \Circle & \Circle & weekly \\
R3 & US & B. Sc.  & Red Teamer & In-house & 11-20   & $<$  1,000 & & \CIRCLE & \CIRCLE & \Circle & & \CIRCLE & \Circle & \LEFTcircle & \Circle & multiple/week \\
M1 & EU & M. Sc.  & CISO & Managed & 6-10   & $>$ 10,000 & & \CIRCLE & \CIRCLE & \CIRCLE & & \CIRCLE & \CIRCLE & \LEFTcircle & \LEFTcircle & daily \\
M2 & US & M. Sc.  & CISO & In-house & 11-20   & $<$  200 & & \CIRCLE & \LEFTcircle & \Circle & & \CIRCLE & \CIRCLE & \Circle & \Circle & weekly \\
M3 & US & Ph. D. & CISO & Managed & 11-20 & $<$  10,000  & & \LEFTcircle & \LEFTcircle & \LEFTcircle & & \LEFTcircle & \LEFTcircle & \LEFTcircle & \LEFTcircle & daily \\
M4 & EU & M. Sc  & CISO & Managed & 2-5 & $<$ 200 & & \CIRCLE & \LEFTcircle & \LEFTcircle & & \CIRCLE & \LEFTcircle & \LEFTcircle & \Circle & daily \\
M5 & EU & M. Sc.  & SOC Manager & In-house & 11-20   & $<$  10,000 & & \CIRCLE & \LEFTcircle & \LEFTcircle & & \CIRCLE & \Circle & \Circle & \Circle & multiple/week \\
M6 & US & M. Sc.  & SOC Manager & In-house & 6-10 & $<$  10,000 & & \CIRCLE & \CIRCLE & \Circle & & \CIRCLE & \LEFTcircle & \CIRCLE & \Circle & daily \\
\bottomrule
\rowcolor{white}
Total & - & - & - & 15I / 10M & - & - & & 19 & 7 & 7 & & 20 & 10  & 5 & \hspace{.2em}2 & multiple/week \\
\bottomrule
\CodeAfter
\MixedRuleShift{9}
\MixedRuleShift{13}
\MixedRuleShift{17}
\end{NiceTabular}
\begin{tablenotes}
\centering
    \item
    \textsuperscript{1} \textbf{Commercial Cloud}: public multi-tenant LLM service (e.g., ChatGPT).
    \quad
    \textsuperscript{2} \textbf{Private Cloud}: dedicated company cloud (e.g., private Azure/AWS).
    \\
    \textsuperscript{3} \textbf{Self-hosted}: LLM operated on company-owned infrastructure.
    \medskip
    \item
    \CIRCLE: Used in a SOC 
    \quad
    \LEFTcircle: Used privately 
    \quad
    \Circle: No usage
\end{tablenotes}
\end{threeparttable}

\end{table*}

To recruit participants with the experience required to address our research questions, we used a recruitment message (see~\Cref{sec:availability}) that described the study as focusing on \glspl{LLM} in \glspl{SOC}. We chose this approach to make our requirements transparent and to increase the likelihood of reaching practitioners with relevant hands-on experience, as several of our research questions asked participants to reflect on their experiences with LLMs.
To ensure we recruited only participants with relevant experience to take part in the study, we defined two acceptance criteria. First, participants had to be practitioners embedded in a \gls{SOC}, i.e., engaged in the day-to-day operations of an identifiable \gls{SOC}. 
Second, participants were required to have a clear understanding of \glspl{LLM} and hands-on experience.
We verified each participant's role and their current SOC engagement by checking their LinkedIn profile. 
We deliberately sampled across the full spectrum of SOC roles including analysts, security engineers, managers, CISOs, and red teamers~\cite{SecurityOperationsCentera} because LLM adoption in a SOC is decided and enacted across this entire hierarchy rather than by any single role.
Both criteria were checked in the demographic screening survey (see~\Cref{sec:availability}). From an initial pool of 29, we excluded one respondent who was not working in a SOC and three who had no prior LLM experience, leaving 25 participants (17 distinct organizations)~(see~\Cref{tab:overview}).

To recruit participants, we leveraged our research team’s industry network (n=10), promoted the study on LinkedIn (n=9), and asked participants to suggest additional contacts (n=6), following the recommendations of prior work~\cite{kaurWhereRecruitSecurity2022}.
We did not provide financial compensation, and participation was entirely voluntary. 
Previous research found that intrinsic motivation is a key driver of creativity~\cite{SelfDeterminationTheoryFacilitation2024,lepperUnderminingChildrensIntrinsic1973,CreativityContextUpdate}, which, due to the wide range of creative brainstorming tasks, was central to our study. Consistent with findings by Serafini et al.~\cite{serafiniEngagingCompanyDevelopers2024}, they expressed a willingness to contribute altruistically while simultaneously gaining insights applicable to their work. Most participants reported that the structured reflection was valuable, as it helped them identify previously unconsidered risks.

\subsection{Data Analysis}
\label{methods:dataanalysis}

We analyzed all interview transcripts using the six-step thematic analysis approach by Braun and Clarke~\cite{clarkeThematicAnalysis2015, braunUsingThematicAnalysis2006}, consistent with established practices in qualitative research~\cite{klemmerUsingAIAssistants2024,wermkeCommittedTrustQualitative2022, groberCloudNotCloud2023,klostermeyerSkippingSecuritySide2024,mcdonaldReliabilityInterraterReliability2019}.
We first familiarized ourselves with the material by conducting the interviews and reading the transcripts (step~1).
All researchers then collectively analyzed an initial set of four transcripts to inductively develop a first codebook (step~2).
Subsequently, the first and second authors independently coded each interview.
Interviews were coded in rounds of three to four. After each round, they merged and reviewed their codes, discussing new codes and resolving disagreements.
We began grouping codes into candidate themes based on their commonalities (step~3).
The codebook and higher-level themes developed iteratively, refined in each round with the insights from newly coded interviews (step~4).
For example, both authors initially coded \textit{LLM-assisted scripting} and \textit{detection rule writing} separately, but later agreed that both reflected security engineering activities and merged them into a single theme. As the coding process progressed, this theme was further refined into the sub-themes \textit{Rule Development} and \textit{Tool Development}.
Throughout the analysis, we repeatedly reviewed the codebook and themes. We continued until each theme captured a clear and distinct central concept and the overall thematic structure was coherent. Between interviews 19 and 25 no substantial new themes or codes emerged, so we stopped recruitment (step~5).
We report the themes, their codes, and example quotes in~\Cref{sec:results} (step~6).
Following Braun and Clarke~\cite{braunOneSizeFits2021}, we do not treat prevalence as evidence of a theme's importance.
However, to better contextualize how often a theme appeared, we use verbal qualifiers (see~\Cref{sec:qualifiers}), as in prior qualitative security studies~\cite{EveryoneThemselvesQualitative,klemmerUsingAIAssistants2024,usmanDistrustBigTech2023,wermkeCommittedTrustQualitative2022}.
In total, we assigned 3,822 codes across 48 unique code categories, yielding a median of 147 per interview.
The final codebook is available online (cf.~\Cref{sec:availability}).
In line with Braun and Clarke's approach, which treats coding as an interpretative, researcher-situated process rather than a consensus-seeking procedure~\cite{braun2022starting, braunReflectingReflexiveThematic2019, braunOneSizeFits2021}, we did not calculate \gls{IRR}, consistent with broader qualitative methodological literature~\cite{mcdonaldReliabilityInterraterReliability2019}.


\subsection{Experience-Based and Envisioned Claims}
\label{met:expVsEnvisioned}
Our exploratory design captured both real-world insights and forward-looking perspectives. 
Consequently, participants' statements could be \textbf{experience-based} if grounded in practice and actual use or \textbf{envisioned} if based on an anticipated use or risk that had not yet materialized.
To avoid conflating current practices and anticipations, we classified each claim as \textbf{experience-based} \CIRCLE\ , if a participant described an LLM use in practice, or \textbf{envisioned} \LEFTcircle\ , if the claim was not validated in practice.
The first author tracked during the interviews and interactive tasks whether participants referred to actual use or envisioned scenarios. If unclear, the first author verified this through a short follow-up question asking whether the participant was referring to an actual use case.
After the interviews, we labeled each claim.
For instance, M1’s SOC operated a \gls{RAG}-based onboarding assistant in production. 
Consequently, we classified the claims made in the context of this use case as \textit{experience-based}.
In contrast, if M1 later discussed a potential risk that had not yet materialized in practice, we classified it as envisioned, even though their assessment was informed by their professional SOC and LLM experience. 
This approach allowed us to analyze the data with a nuanced understanding of the current state of LLM adoption in SOCs and practitioners’ expectations for its future development. 

We apply this distinction per cell in \Cref{tab:llm-challenges-soc-usecases} and \Cref{tab:riskSpecificCountermeasures}.
For example, while describing a Security Engineering use case during flowchart creation, one participant identified interoperability as a challenge, because it was not tied to a system they had operated, we marked it as \textit{envisioned}~\LEFTcircle, whereas grounding the same challenge in a deployed system would have marked it \textit{experience-based}~\CIRCLE.
A table cell is experience-based \CIRCLE\ if at least one participant grounded it in practice, envisioned \LEFTcircle\ if it was only anticipated, and not identified \Circle\ if no participant raised it.

\subsection{Limitations}

First, we report practitioners' perceptions, not objectively validated \gls{LLM} capabilities or limits. 
Therefore, some stated challenges may reflect incomplete \gls{LLM} understanding.
Second, self-reported data is subject to social desirability bias. Participants may have withheld unauthorized \gls{LLM} usage or overstated how thoroughly they validate \gls{LLM} output. We mitigated this by assuring anonymity and stating that their security practices would not be evaluated. 
Third, the study's \gls{LLM} focus and voluntary participation may have self-selected practitioners with greater interest in \glspl{LLM}, and thus may not exactly represent the opinions and experiences of general SOC practitioners. 
Moreover, in the RQ2 task, we ask participants to create an ideal LLM-driven workflow. This implies that LLM integration in such a workflow is desirable and may have encouraged positive responses. 
To counteract overly positive assessments, we included critical engagement with the workflow. We note that the framing may also have surfaced integration opportunities that would not have come up in a more open-ended discussion.
The walkthrough of well-documented LLM-specific risks may also have amplified their salience. Yet participants surfaced concerns freely beforehand and raised adversarial attacks unprompted. We accepted this trade-off so that hype and possible over-reporting would not leave well-documented risks overlooked.
A 90-minute session could not cover all \gls{LLM}-specific risks, we may have missed a risk practitioners would have rated highly.
Finally, despite targeted LinkedIn recruitment, our sample predominantly comprised individuals from Western Europe, reflecting existing networks, and included only four women, mirroring the gender imbalance in cybersecurity~\cite{ShareWomenCybersecurity} and sampling challenges reported in previous studies~\cite{hielscher_lacking_2023, hielscher_employees_2023, danilova_code_2021,alahmadi99FalsePositives2022f,mink_everybodys_2023}.

\section{Results}
\label{sec:results}

\begin{figure}[t]
    \centering
    \includegraphics[width=0.95\linewidth]{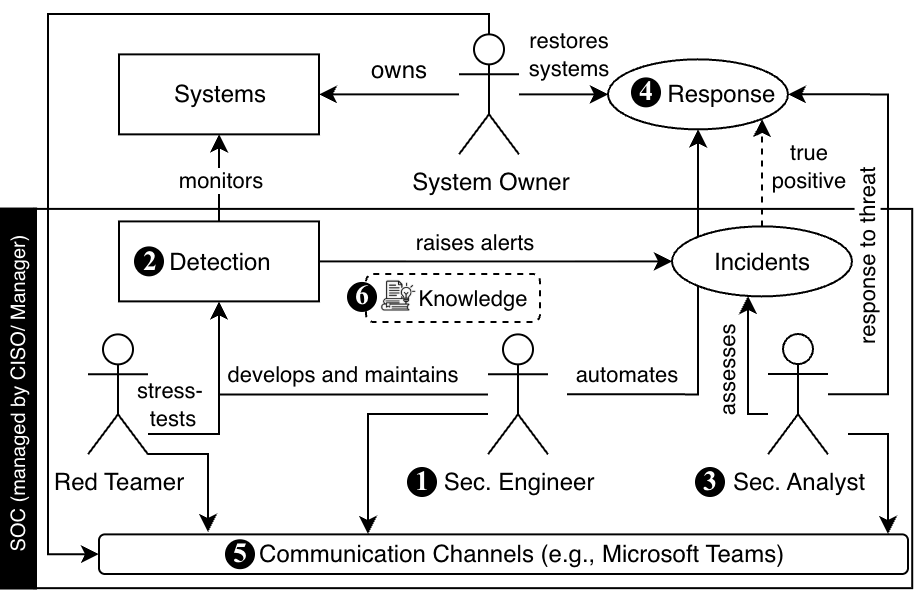}
    \caption{High-level overview of use cases for LLM applications in SOCs. We identified six categories where LLMs may provide operational support: \usecase{1} Security Engineering, \usecase{2} Incident Detection, \usecase{3} Incident Analysis, \usecase{4} Incident Response, \usecase{5} Communication, and \usecase{6} Knowledge. This diagram maps these categories to key SOC roles and workflows.}
    \label{fig:LLMApplicationsInSOCs}
\end{figure}

We outline the findings from our qualitative analysis and address our research questions. We present our findings at the level of themes (see~\Cref{methods:dataanalysis}).
We distinguish experience-based \CIRCLE\ and envisioned \LEFTcircle\ claims (see ~\Cref{met:expVsEnvisioned}).
\Cref{fig:LLMApplicationsInSOCs} provides a high-level overview of the use cases we identified, grouping 15 use cases into six categories and mapping them to key SOC roles and workflows. We detail each category, including its anticipated impact and feasibility, in \Cref{res:rq1}. We then examine the technical and operational challenges practitioners associate with integrating LLMs (\Cref{res:rq2}) and the LLM-specific risks they perceive, together with the countermeasures they propose (\Cref{res:rq3}).

\subsection{Applications, Impact, and Feasibility (RQ1)}
\label{res:rq1}

\begin{figure}[ht]
  \centering
  \includegraphics[width=\linewidth]{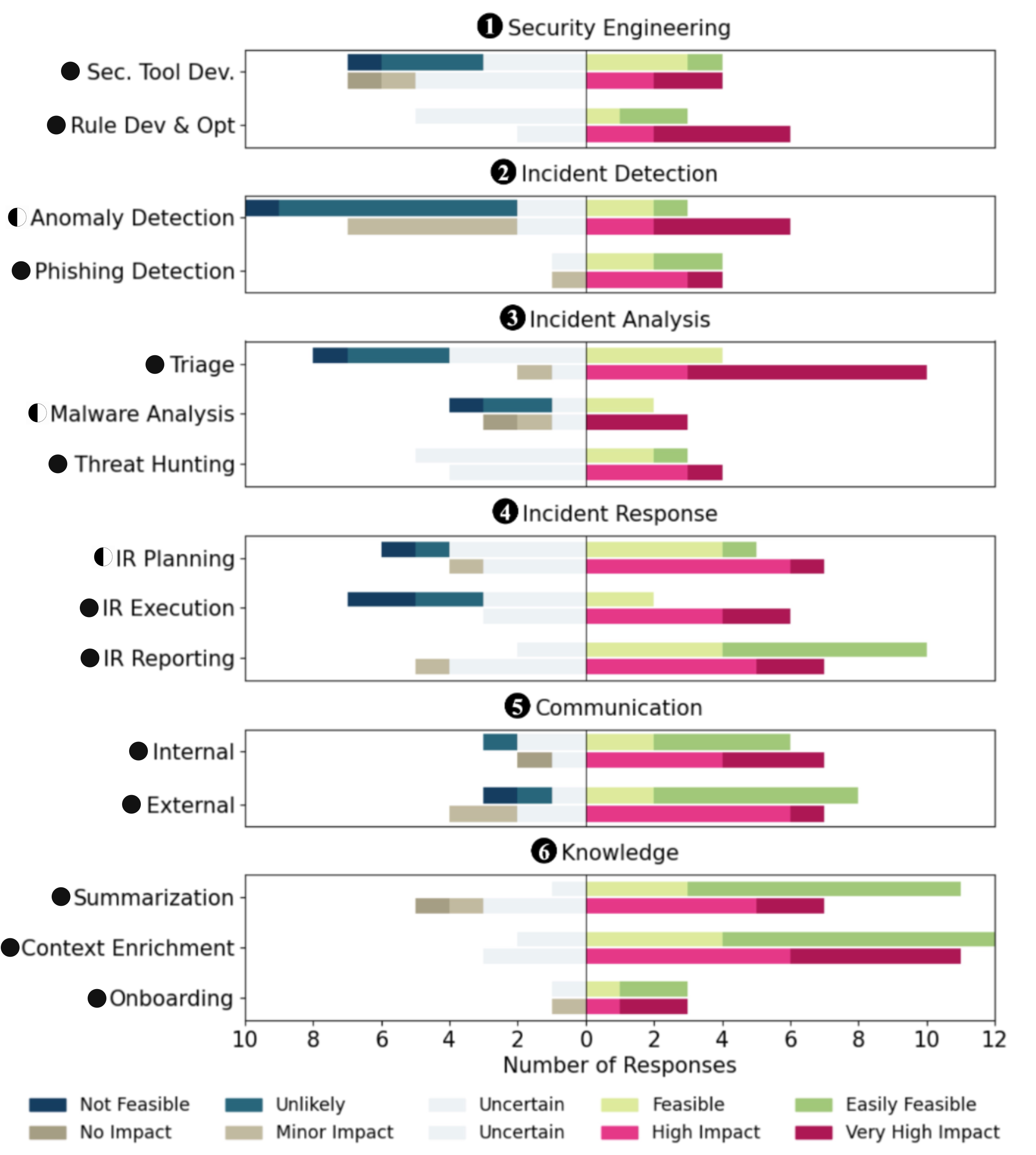}
    \caption{Practitioners' rankings of anticipated impact and feasibility of \gls{LLM} use cases in SOCs, based on participants' brainstorming and ranking activities conducted on the interactive Miro board (see~\cref{fig:ImpactFeasibilityExample}).}
  \label{fig:ImpactAndFeasibilityEvaluation}
\end{figure}

In this section, we discuss all identified use cases for \glspl{LLM} in \glspl{SOC}, along with their perceived impact and feasibility.
\Cref{fig:ImpactFeasibilityExample}
illustrates our interactive think-aloud assessment, and \Cref{fig:ImpactAndFeasibilityEvaluation} provides an overview of the discussed use cases and their anticipated impact and feasibility across the interviews.

\subsubsection{Security Engineering}
\label{res:rq1:securityEngineering}

[Sec. Tool Development \CIRCLE]
Security engineers described LLMs as valuable aids in developing security tools, particularly in SOC environments where engineers regularly interact with unfamiliar tools and technologies. For example, R1 used an LLM to build a custom script to stress-test a Cisco device without first studying its documentation. A majority used these models to \squote{jumpstart} coding efforts even on previously unknown systems.
Yet this same shortcut concerned practitioners. E3 stressed: \squote{You must understand what's happening before running LLM-generated scripts in customer environments.}

[Rule Development \LEFTcircle]
One specific area of security engineering is detection rule development, since poorly designed rules generate excessive false positives, leading to issues such as alert fatigue~\cite{alahmadi99FalsePositives2022f}. 
In particular, many security engineers emphasized that, for this reason, rule creation \eolquote{requires careful fine-tuning.}{E5} 
Moreover, E9 characterized it as \squote{time-consuming} and \squote{labor-intensive,} underscoring the impact on their daily workload. 

[Rule Development \CIRCLE]
Some had already experimented with \glspl{LLM} to save time: \eolquote{I used it to create detection rules, although this didn't always work perfectly [...] the domain-specific knowledge about the products was missing [...] things were mixed up.}{E5} They emphasized that security rule engineering requires in-depth knowledge of the customer's environment, which limits LLMs in \glspl{SOC}.  

\subsubsection{Incident Detection}
\label{res:rq1:incidentDetection}

[Anomaly Detection \LEFTcircle]
About half of the participants envisioned LLMs independently detecting anomalies in log data and triggering alarms. 
However, they raised two major concerns, matching the low-feasibility placements in~\Cref{fig:ImpactAndFeasibilityEvaluation}. First, they argued that alarm generation must not be left to chance and requires a deterministic approach: \eolquote{You cannot roll the dice when generating alarms.}{R2} 
Second, real-time anomaly detection requires processing vast datasets, demanding substantial computational resources:~\eolquote{We generate millions of logs daily, we can't just dump tons of gigabytes in there [LLM].}{M2} 

[Phishing Detection \LEFTcircle]
Participants identified phishing detection as a special case. They agreed that \squote{traditional methods} had been exhausted. They suggested that LLMs could be used to augment these approaches via semantic analysis. 

[Phishing Detection \CIRCLE]
The SOC of A2 built an LLM-based application to analyze over 300 emails that had evaded their spam filters, and reflected on this: \squote{So far, I am pleasantly surprised by the phishing [prevention] skills that these models demonstrate.}

\subsubsection{Incident Analysis}
\label{res:rq1:incidentAnalysis}

[Triage \LEFTcircle] In a SOC, triage is a Tier~1 task in which analysts validate and prioritize alerts by confirming their criticality and filtering false positives~\cite{SecurityOperationsCentera}. In this context, M2 suggested: \squote{LLMs could triage alerts by comparing them to historical data. If an alert matches a previous pattern, the LLM could provide a summary and recommend actions based on past responses.} However, analysts cautioned and anticipated a central limitation: \eolquote{LLMs can't detect alerts they weren't trained on, so it's really just handling the Tier 1, repetitive stuff.}{A7} 
Overall, they considered an LLM-based triage tool, especially when enriched with historical data, as both feasible and valuable. However, a majority clearly preferred assistance over automation.

[Triage \CIRCLE] A3 experimented with LLM-based triage and found: \squote{LLMs are often too generic to effectively address the unique systems of our customers.} 
A limitation that historical data enrichment may help overcome. However, none of our participants reported such a solution.

[Malware Analysis \LEFTcircle] A few participants considered using LLMs for malware analysis tasks such as analyzing obfuscated JavaScript code. They also saw potential for LLMs to assist in behavior analysis within a sandbox environment, noting: \eolquote{We use a Detonation Chamber, where you throw it [malware] in and see what it does. An LLM would be useful, for example, to summarize that.}{A2} 
Beyond this, however, participants considered the actual de-obfuscation task an \squote{unsolvable problem} for LLMs, since they lack understanding of obfuscated code and thus cannot reliably reverse-engineer it.
In the same vein, A4 located the limit in the model rather than its inputs, describing an \gls{LLM} as \squote{a puzzle} that \squote{can puzzle well, but in the end does not know what the motive is.}

[Threat Hunting \CIRCLE]
Threat hunting is a Tier~2 activity in which analysts proactively search for indicators of compromise described in the latest \gls{TI}~\cite{SecurityOperationsCentera}. 
Many participants valued LLMs as a translation layer between analysts and the \gls{SIEM}, generating query syntax (e.g., KQL for Sentinel) from natural language. 
M1 pointed out that vendor-integrated solutions such as Microsoft Security Copilot~\cite{MicrosoftSecurityCopilot} or Splunk AI~\cite{SplunkAI} work well: \squote{My SIEM creates the query for me, which is extremely helpful because I don’t need any technical knowledge}.
Others found that general-purpose LLMs frequently produced incorrect queries. R1 tied this to their lack of domain understanding: \squote{ChatGPT, is like one LLM for the whole world, so it doesn't work very well for security purposes.}

\subsubsection{Incident Response}
\label{res:rq1:incidentResponse}

[IR Planning \LEFTcircle]
\Gls{IR} aims to contain detected security threats before they escalate. This is typically achieved through structured workflows and playbooks that define mitigation steps~\cite{SecurityOperationsCentera} to be taken in certain scenarios. Our participants noted that LLMs could support these structured planning processes, for instance, by analyzing existing playbooks to uncover overlooked risks. However, they stressed that such systems must remain pure supportive tools, with final responsibility resting on humans.

[IR Execution \LEFTcircle]
Analysts presented both advantages and critical arguments regarding the use of LLMs for automation in \gls{IR}.
For example, A4 warned that automated actions are too risky and directly questioned their feasibility, while A1 argued that the risk of acting delayed is greater than the risk of LLM-based misclassification: \squote{If an account is accidentally locked, it's not a big deal, the admin can unlock it the next day.}

[IR Execution \CIRCLE] Our CISOs added another perspective to this discussion. They explained that billions of alerts are triggered daily and that automation is therefore not optional but essential.
For instance, M1, the CISO of an international enterprise with more than 150,000 employees, reported that they had already deployed an LLM-based application to automate parts of \gls{IR}, such as triaging an \squote{impossible traveler} alert, where a single account logs in from geographically distant locations within an implausibly short time.

[IR Reporting \CIRCLE]
As a part of \gls{IR}, SOC teams inform stakeholders about detected threats through structured incident reports~\cite{SecurityOperationsCentera}. Our participants described this process as \squote{tedious}, and E7 clarified: \squote{It is not an analyst's favorite task. Thus, the quality varies.} In addition, E2 explained: \squote{Analysts spend a significant amount of time to fill report templates and write summaries for alerts.} Therefore, some experimented with \gls{LLM}-assisted report writing. E2 continued: \squote{We’re currently testing \glspl{LLM} to automate the text production in reports, but we’re unsure if the quality will be sufficient for our customers.} 
We later received an update from E2, who reported that it is now in production use at their SOC.
Rankings in~\Cref{fig:ImpactAndFeasibilityEvaluation} likewise clustered reporting at high impact and high feasibility.

\subsubsection{Communication}
\label{res:rq1:communication}

[Communication \CIRCLE]
Participants noted that LLMs already enhance communication tasks such as drafting executive summaries, for example by bridging gaps in technical terminology with executives or customers. Most viewed these benefits as modest compared to other applications. 

\subsubsection{Knowledge}
\label{res:rq1:knowledge}

[TI Summarization \CIRCLE]
Practitioners must continuously maintain an overview of the threat landscape, which is a challenging task given the sheer volume of \gls{TI} available. Almost all of our participants reported that LLMs are helpful for summarizing and contextualizing \gls{TI} data.
For instance, M2 pointed out: \squote{There’s a ton of threat intel every day, it's hard to know where to start or what to look for, manually processing it all is impossible [...] summarizing large volumes of threat intelligence [with LLMs] is incredibly useful for day-to-day security operations.}
M3 added: \squote{TI summarization is pretty straightforward. They are language-generation tasks, and that's exactly what these tools [general-purpose LLMs] were built for, and as they save a lot of time, they have impact.} 


[Context Enrichment \CIRCLE]
In a SOC, analysts must assess alerts and logs and correlate them with contextual data such as asset information, historical events, and \gls{TI}~\cite{SecurityOperationsCentera}. 
SOCs deal with many systems that even seniors lack familiarity with. 
For instance, R1 explained: \squote{I got a log of a wireless LAN controller from Cisco. I didn't know the structure and the device, scrolling through it was impossible. So I fed the \gls{LLM} the documentation for the device, and then I entered the log file and asked to process it. [...] That helped a lot because I'm not familiar with these wireless controller logs.} 
This kind of context enrichment was perceived as very useful by many participants across multiple scenarios, including log analysis, alert triage, incident investigation, malware analysis, threat intelligence correlation, and understanding unfamiliar tools, systems, or vendor-specific technologies, in particular, under time pressure.

[Onboarding \CIRCLE] For instance, M1 reported that they had already deployed a RAG-based LLM solution in production to train new employees, especially in their Indian office, where turnover is high. They explained: \squote{Onboarding is now much faster. We even built a badge system to certify analysts' skills. It's cool, it generates training content by itself, classified by skill level and topic, and no one can just memorize old questions.}
E7 reported positively on deploying a knowledge database searchable via an interactive \gls{LLM} as an interface between junior analysts and the SOC's diverse technical systems: \squote{We humans aren’t designed to search and analyze 10,000 pieces of information. That’s where \glspl{LLM} are super helpful because they offer us this human-like way to search through data based on questions and answers ... and if the juniors ask the tool instead of us, then we have more time to take care of more important things.} 

\begin{summarybox}{Summary RQ1}
    Participants not only envisioned but also actively used and deployed LLM-based tools across 15 use cases. Language-centric such as summarization and \gls{IR} reporting clustered at high-impact, high-feasibility (cf.~\Cref{fig:ImpactAndFeasibilityEvaluation}). 
    For more technical or organization-specific tasks, such as query generation and rule writing, many practitioners judged general-purpose models \squote{too generic} and noted missing organization- or product-specific knowledge.
\end{summarybox}

\subsection{Technical and Operational Challenges (RQ2)}
\label{res:rq2}

\begin{table}[t]
\renewcommand{\usecase}[1]{\Circled[fill color=black, inner color=white, inner ysep=2.5pt, inner xsep=2.5pt]{\tiny #1}}

\newcommand{\pie}[1]{%
\hspace{.12em}\begin{tikzpicture}[baseline={(0,-.55ex)}]
 \draw[line width=0.25pt] (0,0) circle (.8ex);
 \pgfmathsetmacro{\absangle}{abs(#1)}
 \ifboolexpr { test {\ifnumcomp{#1}{<}{0}}}{%
  \fill[rotate=90] (.8ex,0) arc (0:{\absangle}:.8ex) -- (0,0) -- cycle;%
 }{%
 \fill (.8ex,0) arc (0:#1:.8ex) -- (0,0) -- cycle;%
 }%
\end{tikzpicture}\hspace{.1em}%
}

  \def\Angle{55}
  \newcommand{\rotheader}[2]{\def\Angle{#1}\Rot{\textbf{#2}}}
    \caption{Technical and operational challenges of LLM integration across SOC use cases.}
    \label{tab:llm-challenges-soc-usecases}
    \begin{threeparttable}
    \centering
    \setlength{\tabcolsep}{5pt}
    \scriptsize
    \begin{NiceTabular}{l@{\hspace{8pt}}w{c}{12pt}llllw{l}{10pt}w{c}{1.05cm}}
      \toprule
      \multicolumn{1}{r}{\rotheader{55}{\textbf{Use Case Categories}}} & \rotheader{55}{\usecase{1} Sec. Engineering} & \rotheader{55}{\usecase{2} Incident Detection} & \rotheader{55}{\usecase{3} Incident Analysis} & \rotheader{55}{\usecase{4} Incident Response} & \rotheader{55}{\usecase{5} Communication} & \rotheader{55}{\usecase{6} Knowledge} & \multicolumn{1}{l}{\rotheader{55}{Overall}} \\
      \midrule
      \textbf{Response Count} & 6 & 5 & 10 & 13 & 8 & 14 & 25 \\
      \midrule
      \rowcolor[gray]{0.9} \textbf{Technical Challenges}  &  &  &  &  &  &  &  \\
      \midrule
      Reliability & \LEFTcircle & \CIRCLE & \LEFTcircle & \LEFTcircle & \LEFTcircle & \CIRCLE & \CIRCLE \\
      Interoperability & \LEFTcircle & \LEFTcircle & \CIRCLE & \LEFTcircle & \LEFTcircle & \CIRCLE & \CIRCLE \\
      Generic Outputs & \CIRCLE & \CIRCLE & \CIRCLE & \CIRCLE & \CIRCLE & \CIRCLE & \CIRCLE \\
      Fine-Tuning & \Circle & \Circle & \Circle & \LEFTcircle & \Circle & \LEFTcircle & \LEFTcircle \\
      Explainability & \Circle & \Circle & \LEFTcircle & \LEFTcircle & \Circle & \LEFTcircle & \LEFTcircle \\
      \midrule
      \rowcolor[gray]{0.9} \textbf{Operational Challenges}  &  &  &  &  &  &  &  \\
      \midrule
      Economic Pressure & \LEFTcircle & \LEFTcircle & \LEFTcircle & \LEFTcircle & \LEFTcircle & \LEFTcircle & \LEFTcircle \\
      AI Governance & \LEFTcircle & \LEFTcircle & \LEFTcircle & \CIRCLE & \CIRCLE & \LEFTcircle & \CIRCLE \\
      Misconceptions & \LEFTcircle & \LEFTcircle & \LEFTcircle & \LEFTcircle & \LEFTcircle & \LEFTcircle & \LEFTcircle \\
      Legal Uncertainty & \LEFTcircle & \LEFTcircle & \CIRCLE & \CIRCLE & \Circle & \LEFTcircle & \CIRCLE \\

      \bottomrule
      \CodeAfter
\MixedRuleShift{2}
\MixedRuleShift{8}
    \end{NiceTabular}

\begin{tablenotes}
  \centering
  \scriptsize
  \item
    \CIRCLE: experience-based \quad
    \LEFTcircle: envisioned \quad
    \Circle: not identified
\end{tablenotes}

\end{threeparttable} 
\end{table}

During each interview, participants sketched a flowchart of their ideal SOC workflow. 
Those who already used LLMs in operations often chose a current workflow rather than a hypothetical redesign.
We then asked to identify technical and operational challenges. In the following we report the technical and operational challenges identified by our practitioners.
Following our data analysis procedure~(\Cref{methods:dataanalysis}), we classified use cases (theme and subtheme) in flowchart notes. Challenges mentioned while discussing a note received a second, independent code. A cell in~\Cref{tab:llm-challenges-soc-usecases} is the co-occurrence of those two codes.
For example, in~\Cref{fig:MiroBoardLLMSOCIntegrationExample} we classified \textit{Triage \& Recommended Actions} as the subtheme Triage under Incident Analysis. While thinking aloud about this node, E5 said the hard part was \squote{making these data sources accessible to LLMs.} We coded that statement as Interoperability. That pair is the Interoperability~$\times$~Incident Analysis cell in~\Cref{tab:llm-challenges-soc-usecases}.

\subsubsection{Technical Challenges}
\label{res:rq2:technicalChallenges}

[Reliability \CIRCLE] Most participants emphasized that the success of LLM integration hinges on the reliability of model outputs. 
In particular, they warned that hallucinations can lead to overlooking genuine threats or interrupt legitimate business activities, thereby \squote{harming reputation} and potentially \eolquote{deterring customers.}{E2}
However, about half of the participants attributed these issues to inadequate contextualization: \eolquote{I think models today lack a lot of context and not only technical but also business priorities.}{M6} In addition, R3 noted: \squote{You can train a perfectly good model, but if you're giving it bad data, then that won't go well.} 
They argued that the \gls{LLM} itself may not be to blame. Instead, they pointed out \eolquote{the inability of the SOC to feed the \gls{LLM} with the correct data,}{E7} questioning whether most SOCs have reached the maturity necessary for effective LLM integration. 

[Interoperability \CIRCLE] E5 stated: \squote{Many customers have bought a variety of security solutions that don’t play well together, aren’t well-configured, or haven’t even been deployed,} underlining persistent interoperability challenges within SOCs. Further, E5 emphasized: \squote{The significant challenge is making these data sources accessible to LLMs.} Our findings clearly show that connecting diverse customer systems and thus adequately contextualizing LLMs is perceived as a significant technical challenge. Interoperability only emerged when participants discussed LLM-based tools, rather than general-purpose manual use (e.g., ChatGPT). 

[Generic Outputs \CIRCLE] Many practitioners complained about generic responses. 
They emphasized that general-purpose models such as ChatGPT and Gemini often lack the necessary technical depth and nuanced understanding of specific customer contexts. 
As R2 put it: \squote{A pure LLM [i.e., an LLM without adequate contextualization] does not know in which environment the alarm occurred. However, this is crucial for determining whether it is a false positive or a true positive. A pure LLM cannot see this, it does not know the corporate environment.} 
M6 illustrated this: \squote{it [the LLM] kept flagging our nightly backup jobs as suspicious. Big encrypted data transfers running at night ticked all the boxes for an exfiltration alert, but of course, these were just our scheduled backups. The LLM didn’t know that because it had no understanding of our environment or normal processes.} 

[Fine-Tuning \LEFTcircle] While participants communicated a demand for more specialized models, a few foresaw another technical challenge in the lack of suitable training data for fine-tuning. In this context, E1 explained that due to the highly specific SOC requirements, each SOC would be obliged to fine-tune models independently in the absence of third-party vendors. 
In this context, R2 made the following statement: \squote{You need a training dataset, and we [SOCs] simply don't have that.} Later, M6 added that SOCs also lack the necessary ML expertise to successfully carry out fine-tuning.


[Explainability \CIRCLE] Moreover, many identified lack of explainability as a key challenge in security operations: \eolquote{We are now at a point where we have moved from ‘I can explain why I got from A to B’ to ‘I got from A to B and that's it’,}{A4} and, particularly in the case of incident assessment, wanted both reasoning and sources in order to verify the assessment: \eolquote{As we operate in a critical environment I want the decision to be reasonable, do fact checking. Does this decision make sense?}{E1}  

\subsubsection{Operational Challenges}
\label{res:rq2:operationalChallenges}

[Economic Pressure \LEFTcircle] Many participants perceived an economic pressure driving the LLM integration into SOCs. For example, E1 predicted: \squote{Those who engage intensively with this topic will have a significant advantage in a few years.} 

[Economic Pressure \CIRCLE] In practice, E7 described the situation in their SOC as follows: \squote{We are already using LLMs in some areas, but we are still in the early stages. The pressure to adopt is immense, and we are trying to keep up with the pace of change. ... The train is moving at full speed, and many companies are investing heavily. There's no turning back now.}

[AI Governance \LEFTcircle] A majority of participants expressed concerns about human governance. In particular, CISOs and Managers questioned who controls the use of LLMs in SOC workflows. M6 warned: \squote{I think the tool is in place, and habits shift. People use it in their day-to-day life, and it works. They start trusting the LLM. They also use it in security operations, and it works fine, so at one point, they skip the review, because it’s easier, right? And then they skip it again and again. Habits shift, and in the end, the LLM makes more and more decisions, and we are not even noticing it.} This statement highlights the risk of delegating decision-making to a proprietary system without clear organizational approval or oversight. They further raised questions about accountability. For instance M2 stated: \squote{If an LLM makes a mistake, who is responsible? The analyst? The SOC manager? The CISO? The vendor?} and continued \squote{If you have to explain to a customer that they have been encrypted [by ransomware] because an LLM made a mistake, how do you think they will react? They won't care about fancy tech. They want accountability.}

[Misconceptions \CIRCLE] We found that many security practitioners acknowledged they do not understand LLM capabilities. They stressed that these systems act as \eolquote{black boxes,}{A2} making it nearly impossible to trace how decisions are reached.
For instance, E6 admitted, \squote{I don't fully understand LLMs. I'm not really sure what they do in the background, even though I use them daily for various tasks.} 
R1 added: \squote{It's not just non-technical people; even technical people struggle with them. They expect it to do things it can't.} 
[Misconceptions \LEFTcircle]
According to M2, such misconceptions \squote{can lead to a false sense of security}, where blind trust in technology replaces critical thinking and expertise, which our practitioners described as particularly dangerous in security operations, highlighting that SOCs are handling \squote{sensitive customer logs and systems.} 

[Legal Uncertainty \LEFTcircle] We noticed a surprising number of contradictions regarding compliance among our EU practitioners.
For instance, A4 (EU) was convinced that there were no regulations \squote{at all,} and demanded that \squote{lawyers have to speak up,} while E2 (EU) thought that they were \squote{showstoppers,} especially for \gls{LLM}-driven automation, with R1 (EU) adding: \squote{I think law enforcement is not ready for what's happening.}
In contrast, our practitioners based in the US expressed far less concern about legal restrictions.

\begin{summarybox}{Summary RQ2}
    Practitioners located the core technical challenge not in the LLMs themselves but in the SOC, attributing unreliable outputs to inadequate contextualization stemming from interoperability issues. They framed LLM readiness as a question of organizational maturity that many SOCs perceive themselves to lack. At the same time, many participants described intense economic pressure driving adoption while operational challenges such as legal uncertainty and governance remain unresolved.
\end{summarybox}

\subsection{Risk Perceptions and Countermeasures (RQ3)}
\label{res:rq3}

In this section, we report on the final interview segment, in which participants evaluated LLM-related risks in SOCs and discussed potential mitigation strategies. 

\begin{figure}
    \centering
    \includegraphics[width=\linewidth]{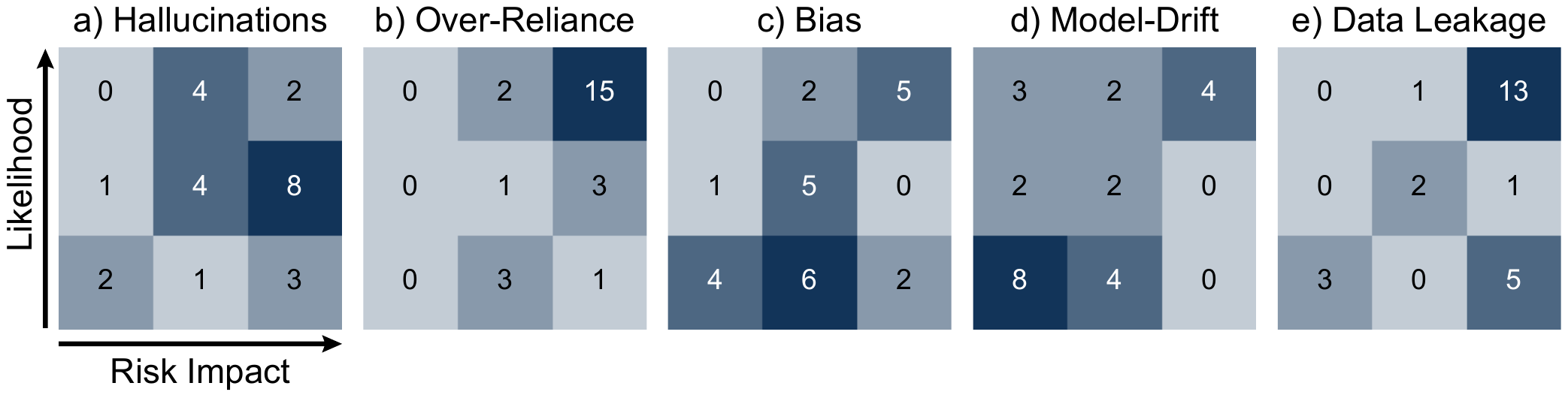}
    \caption{Practitioners' risk assessments of selected LLM-specific risks (cell values are response counts), based on the corresponding interactive Miro-board activity (see~\cref{fig:MiroBoardRiskAssessmentExample}).}
    \label{fig:riskAssessment}
\end{figure}

\subsubsection{Risk Perceptions}
\label{res:rq3:riskPerceptions}

\Cref{fig:riskAssessment} shows practitioners' assessments of predefined \gls{LLM}-specific risks during the interactive Miro-board activity (cf.~\cref{met:interviewguide}).

[Hallucinations \LEFTcircle] A majority of participants perceived the impact of hallucinations as critical, particularly because of the potential to mislead decision making: \eolquote{If it’s wrong and I handle an incident incorrectly, it might not be contained properly, leaving attackers active in the network}{A3}. M6 added: \squote{If it wrongly escalates, triggers a containment action, like blocking a legitimate partner domain or isolating a critical server, it disrupts business.} 
It is worth highlighting that many participants did not attribute hallucinations to the model itself: \eolquote{It's pure mathematics, if the data set is so sparse that 1+1 equals 3, the machine will present it as fact.}{E7} Instead, about half attributed hallucinations to a lack of context regarding technical details, customer processes, and systems (cf.~\Cref{res:rq2:technicalChallenges}). 

[Hallucinations \CIRCLE] Some observed that LLMs will \squote{always} hallucinate when confronted with unseen, sophisticated attacks. E9 explained, \squote{When an APT group attacks, it is not possible for the LLM to detect them because they use their own tools that no one has seen before [...] and there is no data that the LLMs could use to identify these attacks.} 
They suggested that LLMs struggle to detect threats beyond their training data or contextual inputs. 
Interestingly, some practitioners argued that only experts can reliably spot errors in \gls{LLM} output, which conflicts with the view that LLMs are most helpful for junior staff.
Overall, the fear of missing critical alerts, especially in decision-making tasks, drove a strong preference among most participants for assistance systems.

[Over-Reliance \CIRCLE] Overall, over-reliance received the highest risk rating across almost all interviews, primarily because it could lead to false negatives, which participants considered particularly hazardous in SOC environments:
\eolquote{In a SOC, there’s no safe environment to see what happens. If the model gets something wrong and an analyst blindly trusts the response, that could mean a breach slips through, or we waste resources ... You don’t get a second chance to patch the mistake.}{M3}
Most attributed over-reliance to human factors rather than technical shortcomings. A few practitioners admitted to not always verifying LLM outputs, and attributed this behavior to human factors, which we summarize in \Cref{tab:overRelianceHumanFactors}.

\renewcommand{\arraystretch}{1.3}
\begin{table}[t]
  \centering
  \caption{Human factors that contribute to over‑reliance on large language models in SOCs according to our participants.}
  \label{tab:overRelianceHumanFactors}
    \footnotesize
  \begin{tabularx}{\linewidth}{p{2.1cm} X}
    \toprule
    \textbf{Human factor} & \textbf{Explanation} \\
    \midrule
    \rowcolor[gray]{0.95}
    Lack of expertise & Analysts with limited domain knowledge tend to trust automated suggestions more than seniors. \\
    Familiarization & Constant use over time reduces the ability to perform independent checks. \\
    \rowcolor[gray]{0.95}
    Normalization & As trust increases, behavior gradually drifts toward uncritical acceptance. \\
    Time pressure & Time pressure and heavy workloads encourage SOC analysts to accept LLM outputs as a shortcut. \\
    \rowcolor[gray]{0.95}
    Burnout \& fatigue & Mental exhaustion reduces cognitive resources for critical evaluation. \\
    Presentation bias & LLMs can produce logically coherent but incorrect explanations, even when wrong. \\
    \rowcolor[gray]{0.95}
    Identification & When decision‑making is shared with an AI system, analysts feel less personally accountable. \\
    \bottomrule
  \end{tabularx}
\end{table}

[Over-Reliance \LEFTcircle] Interestingly, a majority of practitioners argued that LLMs should lean toward false positives. R2 captured the reasoning: \squote{If something goes wrong, it can go seriously wrong. Therefore, \glspl{LLM} should probably lean toward false positives, leaving the final verification to humans.}
R1 echoed: \squote{When an \gls{LLM} is involved in security decision-making, it's better to overreact.} 
Moreover, A4 anticipated an erosion of skills as analysts grow accustomed to \glspl{LLM}, asking, \squote{How many people can still read a conventional map nowadays?} A4 warned that this problem would become \squote{exponentially worse over time,} eventually leading to a SOC that \squote{raises a generation that relies on it [LLMs] 100\%} and can no longer function without them.

[Bias \LEFTcircle] Bias was among the risks practitioners engaged with least, and their ratings diverged sharply, tracking whether they expected it to cause false negatives. The clearest concern was detection, where E6 warned that a model assisting alert prioritization might favor frequent threats and overlook rarer, more sophisticated attacks. Yet E6 still felt unable to act, since \eolquote{it's really not up to us, we're talking about the training data and the company providing the model.}{E6} Many dismissed bias outright, as M3 put it, a \squote{technical system is not biased, unless it [bias] is implemented.}

[Bias \CIRCLE] In practice, participants observed bias mainly in phishing detection, where broken English or foreign-language patterns led the model to flag messages as malicious. For other tasks such as \gls{TI} summarization, they doubted bias mattered at all.

[Model-Drift \LEFTcircle] Almost all participants linked model-drift to the evolving threat landscape. E2, for instance, explained: \squote{It's critical, as the situation in security is constantly changing.} Interestingly, some contradicted this: \eolquote{When I look at the types of attacks that we have faced over the last five to ten years, they are still largely the same}{M2}. Overall, practitioners considered the risk manageable.

\label{res:rq3:dataLeakage}
[Data Leakage \LEFTcircle] The impact of data leakage was perceived as high, because of the sensitive data handled in SOCs (e.g., patient records, financial transactions):~\eolquote{Most of the data we deal with in a SOC is highly critical, customer logs, incidents, sensitive systems. [...] the criticality of the environment demands people to be much more cautious}{A7}. All but one of our practitioners distrusted LLM providers; some were concerned about misconfigurations. For instance, M1 drew parallels to incidents such as Snowflake’s unprotected S3 buckets~\cite{SnowflakeDataBreach} and warned that \squote{the same might happen with OpenAI tenants.} They anticipated that such leaks would significantly \squote{harm [their] reputation} and \eolquote{deter customers.}{E2} 
However, almost all agreed that the risk is easily mitigated by hosting \glspl{LLM} on-premises.

[Data Leakage \CIRCLE] In particular, managers worried about analyst carelessness and data protection violations. M1 recounted: \squote{Employees even forward data to private email to input it into ChatGPT} and bypass organizational safeguards. Moreover, engineers expressed concerns about the risk of data leakage between customer environments. 

[Adversarial Attacks \CIRCLE] Although adversarial attacks were not a separate item in our predefined risk list, we later decided to report them separately because many raised diverse attack vectors, such as prompt injection and data poisoning, unprompted, and expressed strong concerns about them. 
Red team members were especially concerned about LLM tools granted powerful permissions.
R2 found a Copilot misconfiguration that allowed them to access sensitive information, including salary data for all employees.
R1 observed that LLMs are becoming a primary attack target: \squote{Most data now goes through an LLM, so that's where hackers go.}

[Adversarial Attacks \LEFTcircle] 
Participants also cited demonstrations of prompt injections and data poisoning. Through prompt injection, 
A2 warned, \squote{you can completely take over individual LLMs and even inject malicious code,} emphasizing that no reliable defense exists yet.
E8 made the risk conditional on deployment, assigning a high risk to publicly accessible LLMs while rating risk for private applications lower. 
R3 cautioned that on-premises hosting does not eliminate the threat, since \squote{even if I have an LLM on an on-prem server, it is still a target for attackers,} and asked for SOC-specific preparation, such as \squote{incident response playbooks for LLM attacks}.

\subsubsection{Countermeasures}
\label{res:rq3:countermeasures}

\Cref{tab:riskSpecificCountermeasures} in~\Cref{sec:countermeasures} maps the mentioned countermeasures to the discussed risks.

[Input Optimization \CIRCLE] A few reached for prompt engineering as an immediate countermeasure. For instance, E6 reported that hallucinations became less frequent once they decomposed tasks, \squote{I'm not bombarding it with a lot of data and questions [...] I'm giving it very specific, small tasks,} and emphasized the importance of prompt engineering. 
Most likewise perceived context enrichment as an effective way to reduce hallucinations. 
Some also perceived this as an indirect safeguard against over-reliance. By reducing errors in the first place, they reasoned that analysts' reliance would be less consequential. As E3 put it, \squote{if the quality of the LLMs is so good that errors never occur, over-reliance plays no role.}
[Input Optimization \LEFTcircle] Looking ahead, participants envisioned \gls{RAG} as the natural way to ground outputs in data and thereby mitigating hallucinations. However, while \gls{RAG} was invoked in almost all interviews, only one participant mentioned using it in practice.

[Output Verification \CIRCLE] Consistent with their preference for assistive over autonomous systems, most participants emphasized verifying \gls{LLM} output before it takes effect. They described keeping a human in the loop as standing practice. 
E8 explained that \squote{experienced colleagues review the LLM's output and manually cross-check it.}

[Output Verification \LEFTcircle] Moreover, practitioners invoked established principle: \eolquote{One person creates a playbook, and only after the four-eyes principle, when at least two say it's okay, do we officially release it.}{A3} E5 extended the same logic to reject full automation: \squote{an analyst should review the assessment [...] as of today I don't see fully automated handling.} 
Others envisioned \gls{LLM} cross-checks, using a second model to verify the first. In this context, E6 described an agentic setup in which one agent is responsible for highlighting the other's mistakes.

[Model Management \LEFTcircle] Participants envisioned training and fine-tuning as a universal countermeasure. E4, for instance, imagined fine-tuning a model to reliably \squote{recognize registry keys or other IOCs.} In practice, however, they conceded they lacked the data, resources, or expertise to do so. A3 dismissed self-training as \squote{disproportionate,} noting \squote{we can't go and train such a model for this task ourselves}, leaving market offerings as the only realistic, \squote{plug-and-play} option. 
[Model Management \CIRCLE] Practitioners reported always using up-to-date models and described routinely tracking the best-performing models as part of their workflow.

[Data Governance \CIRCLE] Participants viewed measures to prevent data breaches as the most feasible, primarily because they build on practices that SOCs already employ. In this context, M4 noted that each customer \squote{gets their own server with their own customer instance [...] because they all have to be physically separated,} and that the \gls{LLM} tooling simply inherited this separation. 
[Data Governance \LEFTcircle] E1 similarly insisted that mixing customer data \squote{should [...] at its core not even be possible,} treating strict separation as an architectural default.
They further suggested on-premises or private tenant hosting of LLMs, which we categorized under the theme of secure LLM deployment. 
M3 considered it \squote{daily business [...] the same [as] for the SIEM and SOAR} and concluded data leakage was \squote{not a big risk when self-hosting or running models locally.} 
Interestingly, this contrasts with the fact that 19 practitioners in our demographic survey also reported using commercial cloud environments in SOCs. 

[Explainability \LEFTcircle] Explainability was both a challenge (\Cref{res:rq2:technicalChallenges}) and a proposed countermeasure. Many wanted reasoning and sources so they could verify outputs and avoid blind trust.

[Explainability \CIRCLE] Yet others questioned this. A4 cautioned that \squote{source citations are not always reliable} and E5 observed that LLMs can provide sound reasoning even when wrong, suggesting explanations can increase trust in incorrect output rather than mitigate over-reliance.

[Organizational Measures \CIRCLE] Finally, most participants pointed to organizational measures. 
M4 explained that their policy limits the data that may be entered into LLMs and prohibits the use of uncontrolled AI outputs.
Practitioners stressed the importance of clear organizational guidelines, as R1 put it: \squote{we [SOC practitioners] are good at technical problems, but we are bad at solving human problems [i.e., over-reliance].}

[Organizational Measures \LEFTcircle] Moreover, Participants suggested regular awareness training and CI/CD-style validation pipelines as safeguards against over-reliance.

\begin{summarybox}{Summary RQ3}
Most practitioners located the source of failure outside the model, attributing over-reliance to human factors, hallucinations to inadequate contextualization, and bias to people. Their dominant fear was missing a genuine threat, and some warned that LLMs would always fail on novel attacks. 
Many viewed the LLM as a new privileged attack target, while bias and model drift drew little engagement. 
A majority described a gap between envisioned and actual safeguards. In particular, technical safeguards such as RAG and fine-tuning remained mostly envisioned.
\end{summarybox}

\section{Discussion}

\subsection{SOC Readiness, Not Model Capabilities}
\label{dis:readiness}

When our participants reasoned about the limits of \glspl{LLM} in SOCs, they located the central issues more in SOC maturity than in model capability. Hallucinated outputs were attributed to data problems, and biases to human error. About half grounded this in \eolquote{the inability of the SOC to feed the \gls{LLM} with the correct data}{E7 \CIRCLE} (cf.~\cref{res:rq2:technicalChallenges}). Many also pointed to interoperability issues consistent with prior work~\cite{vielberth_security_2020,nepal_burnout_2024}.
While these readiness challenges are not unique to SOCs, they are particularly consequential in security operations, where the cost of errors can be high (e.g., missed intrusions).
However, as SOC experts rather than \gls{LLM} experts, they may misattribute the causes of some failures~(cf.~\cref{res:rq2:operationalChallenges}).
Moreover, findings may age differently. Organizational challenges, such as data readiness in SOCs, may persist longer than limitations tied to current model capabilities. The same applies to mitigations such as \gls{RAG}. As research in this area advances rapidly, newer or more effective countermeasures may emerge.


\begin{researchopportunity}
Future work could identify readiness factors that inform practitioners whether their SOC is mature enough for a given \gls{LLM} use case, rather than assessing model capability in isolation.
\end{researchopportunity}

\subsection{Adoption Despite Concerns}
\label{dis:adoption}

Previous studies have reported low adoption due to concerns about the reliability of AI in SOCs~\cite{mink_everybodys_2023,oesch_assessment_2020,alahmadi99FalsePositives2022f}. Our participants echoed these concerns. They ranked over-reliance above hallucinations, while bias and model-drift drew little engagement~(cf.~\cref{fig:riskAssessment,res:rq3:riskPerceptions}), and largely insisted on keeping a human in the loop, with only a minority willing to trust LLMs to make autonomous decisions~(cf.~\cref{res:rq3:countermeasures}).
These concerns reflect the asymmetric costs in cybersecurity: a false positive can be manually reviewed and discarded, whereas a missed attack is far more expensive.
E1 (\LEFTcircle) pushed this logic further, arguing that \glspl{LLM} should be fine-tuned to \squote{tend more towards false positives} (cf.~\cref{res:rq3:countermeasures}). 
As practitioners delegate security decisions to LLMs, their major concern moves toward the false negative an over-relied-upon model might wave through, away from the false positive concerns in previous SOC work~\cite{nepal_burnout_2024,alahmadi99FalsePositives2022f}. As M1 captured it, \squote{We went from chasing ghosts to missed attacks.}

Yet while these concerns previously led to low adoption and outright rejection of AI systems, many of our participants expressed a strong willingness to adopt them anyway, citing competitive pressure. They argued that defenders must keep pace with attackers, which leaves organizations little choice but to adopt LLMs despite unresolved reliability concerns.
Indeed, a growing number of reports describe cyberattacks facilitated by LLMs~\cite{holleyZerodaysAreNumbered,lengArupLost,mingFakeMilitaryIDs,robins-earlyCEOWorldsBiggest2024,ClaudeMythosPreview} and organizations are rapidly integrating them~\cite{StateAIGlobal,GenerativeAIShowsRapidGrowth2025}.
At the same time, participants noted that \glspl{LLM} embedded in SOC workflows become a highly privileged attack target, adding to the existing challenges~(cf.~\cref{res:rq3:riskPerceptions}).
We are concerned that competitive and economic pressure may accelerate \gls{LLM} adoption before adequate safeguards are in place.


\begin{researchopportunity}
Over-reliance warrants further investigation. Future work could compare senior and junior analysts triaging alerts with and without \gls{LLM} assistance under time pressure, varying the model's outputs, including reasoning, source citations, or confidence indicators, to identify factors that foster over-reliance and the safeguards that mitigate it.
\end{researchopportunity}

\subsection{Envisioned Versus Actual Use}
\label{dis:envisioned-use}

We found a disconnect between envisioned and actual \gls{LLM} use. Some practitioners deploy LLMs in production for language-centric tasks (e.g., report automation, onboarding), while rating high-impact tasks such as incident analysis as low in feasibility (cf.~\cref{fig:ImpactAndFeasibilityEvaluation}). 
Based on their experience with current general-purpose LLMs, many participants judged these models insufficient for use cases requiring specialized security expertise or organization-specific knowledge (\CIRCLE).
This amplified concerns about over-reliance and the fear of missing genuine threats.  
They called for security-specific solutions (\LEFTcircle), citing vendor-integrated threat hunting as a positive example. 

Singh et al.~\cite{singhLLMsSOCEmpirical2025a} analyzed 3{,}090 \gls{LLM} queries from 45 analysts in a single SOC over 10 months, finding that usage was dominated by on-demand sensemaking and context building rather than high-stakes security decisions. Notably, 93\% of queries aligned with established cybersecurity competencies, suggesting analysts are already directing general-purpose LLMs toward security-specific tasks.
However, it remains unclear whether the pattern generalizes across SOCs.
If so, it suggests that using general-purpose LLMs for security-specific tasks is a consistent pattern, reinforcing the demand for purpose-built solutions.


\begin{researchopportunity}
Future work should develop security-specific (e.g., RAG-based or fine-tuned) SOC models and agents on contextual data such as logs, \gls{TI}, and historical incidents and benchmark them against general-purpose baselines. 
\end{researchopportunity}

\subsection{Envisioned Versus Actual Safeguards}
\label{dis:envisioned-safeguards}

A recurring theme across our findings is the disconnect between the safeguards practitioners advocated and those they actually deployed (cf.~\cref{res:rq3:countermeasures}).
For instance, \gls{RAG} was widely envisioned as a mitigation against hallucinations and over-reliance, cited by nearly all participants, yet only one participant reported actual use.
Similarly, fine-tuning was proposed for the same purpose but not implemented by any participants.
This pattern extends to data governance: despite recommending to use on-premises hosting as a safeguard against data leakage, the same participants reported relying on commercial cloud \glspl{LLM} (cf.~\Cref{tab:overview}).
A similar gap appeared in governance: participants envisioned AI governance and policies, yet M1 reported that employees forwarded data to private email accounts to access ChatGPT and bypass these safeguards (cf.~\cref{res:rq3:dataLeakage}).
These countermeasures (human oversight, \gls{RAG}, private hosting, policies) overlap advice already common in human--AI literature; what this sample adds is the gap between advocacy and deployment.
Likewise, while practitioners anticipated that model-provided reasoning and citations would build trust and reduce over-reliance, E5 (\CIRCLE) warned that an \gls{LLM} \squote{can provide a persuasive argument even when it is incorrect,} undermining the assurances these features are meant to offer. Taken together, these observations raise the question of which proposed mitigations translate into meaningful protection under real operational conditions.

In this context, our results also provide a first indication of how policies, roles, and accountability may interact. Policies intended as safeguards were circumvented in practice~(cf.~\cref{res:rq3}). Managers emphasized the value of automation, while analysts stressed the need for manual verification~(cf.~\cref{res:rq1}). Red teamers were especially concerned about LLM tools granted powerful permissions~(cf.~\cref{res:rq3:riskPerceptions}).
However, our results and research design do not provide the data to go deeper into these interactions between policies, roles, and accountability. The study was designed around use cases, challenges, and risk perceptions, not around comparing roles or deployment maturity.

\begin{researchopportunity}
Future work should examine LLM adoption in practice, including how policies, roles, and accountability shape its use and how countermeasures and safeguards perform across the use cases identified in this work. For example, field studies could deploy \glspl{LLM} in real-world SOC workflows to evaluate these factors and identify which safeguards hold in practice.
\end{researchopportunity}

\section{Conclusion}

We conducted a qualitative study with 25 SOC practitioners who had prior LLM experience, combining semi-structured interviews with interactive brainstorming and visualization tasks.
Across the 15 use cases we discussed, many practitioners integrate LLMs for language-centric tasks such as report automation, but perceive high-impact tasks such as incident analysis as low in feasibility.
They judged current general-purpose models insufficient for work that requires security- and organization-specific knowledge. They attributed these shortcomings, as they saw them, less to the models than to the readiness of their SOCs and to human factors.
Despite concerns, many practitioners expressed a strong willingness to adopt LLMs, citing competitive pressure. The safeguards they recommended (RAG, fine-tuning, and on-premises deployment) were widely advocated but rarely deployed.
We interpret this as LLMs entering SOCs not because reliability concerns have been resolved, but because many practitioners see few alternatives.
They described a shift in concern from false positives towards false negatives as they delegate security decisions to LLMs, driven by over-reliance and the fear of missing genuine threats.

\section*{Acknowledgments}
\label{sec:acknowledgments}

We thank all participants for their valuable contribution to this study. Their insights and time were greatly appreciated. We hope that this work will assist both research and operational security in improving their understanding of LLMs in the SOCs. Last but not least, we thank the anonymous reviewers for their valuable feedback.

Jonas Thurner was supported by the research project "Cyber Security Incident Response für KMUs (CySIRK)" (13FH101KB1) of the German Federal Ministry of Education and Research (BMBF) and the nicos AG.
Lea Gröber conducted this work while supported by a fellowship within the IFI program of DAAD, funded by the German Federal Ministry of Education and Research (BMBF).
Funded by the Deutsche Forschungsgemeinschaft (DFG, German Research Foundation) under Germany's Excellence Strategy - EXC 2092 CASA - 390781972.
This research work was supported by the National Research
Center for Applied Cybersecurity ATHENE. This
research was also supported by the Graduate School for
Applied Research in North Rhine-Westphalia (Graduate
School NRW).

\bibliographystyle{IEEEtran}
\bibliography{references}

\appendices
\crefalias{section}{appendix}
\section{Open Science}
\label{sec:availability}

Our replication package, available at \url{https://osf.io/k24j7/overview?view_only=12c9ebe2905d4e2c9647d68024ba99cb}, includes (1) the interview guideline, (2) our recruitment material and written consent form, (3) the demographics questionnaire, and (4) our codebook. To protect participant privacy, we do not share interview transcripts.

\section{Qualifiers}
\label{sec:qualifiers}

\begin{figure}[thb]
    \centering
    \includegraphics[width=0.95\linewidth]{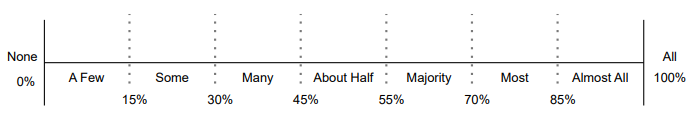}
    \caption{Qualifiers and their respective percentages as used to report our qualitative results. Graphic adapted from Amft et al.~\cite{EveryoneThemselvesQualitative}.}
    \label{fig:quantifiers}
\end{figure}

As is common in thematic analysis, we use qualifiers (see~\Cref{fig:quantifiers}) to better contextualize our findings, similar to previous studies \cite{EveryoneThemselvesQualitative,klemmerUsingAIAssistants2024,usmanDistrustBigTech2023,wermkeCommittedTrustQualitative2022}. We used qualifiers to indicate the frequency with which certain views or concerns appeared across interviews and do not treat this as a measure of statistical significance or importance. This helps distinguish widely shared patterns from individual perspectives. We believe such qualifiers provide useful context on the relative weight of findings without suggesting population-level prevalence.

\section{Countermeasures Mapped to Risks}
\label{sec:countermeasures}

\Cref{tab:riskSpecificCountermeasures} maps each mentioned countermeasure to the risks discussed in~\Cref{res:rq3}, distinguishing experience-based from envisioned claims as described in~\Cref{met:expVsEnvisioned}.

\begin{table}[htb]
  \def\Angle{55}
  \newcommand{\rotheader}[2]{\def\Angle{#1}\Rot{\textbf{#2}}}
    \caption{Mentioned countermeasures mapped to discussed risks.}
    \label{tab:riskSpecificCountermeasures}
    \begin{threeparttable}
    \centering
    \setlength{\tabcolsep}{5pt}
    \scriptsize
    \begin{NiceTabular}{l@{\hspace{18pt}}lllll@{\hspace{12pt}}w{c}{0.85cm}@{\hspace{4pt}}}
      \toprule
      \multicolumn{1}{r}{\rotheader{55}{\textbf{Risks}}} & \rotheader{55}{Hallucinations} & \rotheader{55}{Over-Reliance} & \rotheader{55}{Bias} & \rotheader{55}{Model-Drift} & \rotheader{55}{Data Leakage} & \multicolumn{1}{l}{\rotheader{55}{Overall}} \\
      \midrule
      \rowcolor[gray]{0.9} \textbf{Input Optimization} &  &  &  &  &  &  \\
      \midrule
      Prompt Engineering & \CIRCLE & \Circle & \Circle & \Circle & \Circle & \CIRCLE \\
      RAG & \LEFTcircle & \LEFTcircle & \Circle & \LEFTcircle & \Circle & \LEFTcircle \\
      Context Enrichment & \CIRCLE & \CIRCLE & \LEFTcircle & \LEFTcircle & \Circle & \CIRCLE \\
      \midrule
      \rowcolor[gray]{0.9} \textbf{Output Verification} &  &  &  &  &  &  \\
      \midrule
      HITL & \CIRCLE & \CIRCLE & \LEFTcircle & \LEFTcircle & \Circle & \CIRCLE \\
      4-Eye Principle & \CIRCLE & \CIRCLE & \LEFTcircle & \Circle & \Circle & \CIRCLE \\
      LLM-Crosschecks & \LEFTcircle & \LEFTcircle & \Circle & \Circle & \Circle & \LEFTcircle \\
      \midrule
      \rowcolor[gray]{0.9} \textbf{Model Management} &  &  &  &  &  &  \\
      \midrule
      Training \& Fine-Tuning & \LEFTcircle & \LEFTcircle & \LEFTcircle & \LEFTcircle & \Circle & \LEFTcircle \\
      Model Updates & \Circle & \Circle & \Circle & \CIRCLE & \Circle & \CIRCLE \\
      \midrule
      \rowcolor[gray]{0.9} \textbf{Data Government} &  &  &  &  &  &  \\
      \midrule
      Data Segmentation & \Circle & \Circle & \Circle & \Circle & \CIRCLE & \CIRCLE \\
      Secure Deployment & \Circle & \Circle & \Circle & \Circle & \CIRCLE & \CIRCLE \\
      \midrule
      \rowcolor[gray]{0.9} \textbf{Explainability} &  &  &  &  &  &  \\
      \midrule
      Citation & \LEFTcircle & \LEFTcircle & \LEFTcircle & \Circle & \Circle & \LEFTcircle \\
      Reasoning & \CIRCLE & \CIRCLE & \LEFTcircle & \Circle & \Circle & \CIRCLE \\
      Confidence Indicator & \LEFTcircle & \LEFTcircle & \Circle & \Circle & \Circle & \LEFTcircle \\
      \midrule
      \rowcolor[gray]{0.9} \textbf{Organizational Measures} &  &  &  &  &  &  \\
      \midrule
      Internal Policies & \CIRCLE & \CIRCLE & \Circle & \Circle & \CIRCLE & \CIRCLE \\
      Awareness Trainings & \LEFTcircle & \LEFTcircle & \Circle & \Circle & \LEFTcircle & \LEFTcircle \\
      CI/CD Integration & \LEFTcircle & \LEFTcircle & \Circle & \LEFTcircle & \Circle & \LEFTcircle \\
      \bottomrule
      \CodeAfter
\MixedRuleShift{2}
\MixedRuleShift{7}
    \end{NiceTabular}

\begin{tablenotes}
  \centering
  \scriptsize
  \item
    \CIRCLE: experience-based \quad
    \LEFTcircle: envisioned \quad
    \Circle: not identified
\end{tablenotes}

\end{threeparttable}
\end{table}

\section{Ethics Considerations}
\label{sec:ethics}

We proactively considered the ethical and legal implications for all stakeholders potentially affected by this study. We adhered to the ethical principles outlined in the Menlo Report~\cite{TheMenloReport} and handled all study data and participant information in accordance with the \gls{GDPR}~\cite{GeneralDataProtectiona}. In the following, we discuss the potential risks and benefits for the identified stakeholder groups, as well as the measures implemented to protect them throughout the study. At the time the study was conducted, our institution did not maintain an \gls{IRB}.

\textit{Recruitment Strategy} Our recruitment strategy (see \cref{methods:participants}) allowed us to invite practitioners who actively work in or closely collaborate with \glspl{SOC}. Alternative recruitment channels, such as social media platforms, were avoided, as they would have limited our ability to verify participants’ professional roles. Participation was voluntary and unpaid. This decision was made to avoid financial incentives. Prior work indicates that intrinsic motivation plays an important role in creative and reflective tasks, which were central to this study~\cite{SelfDeterminationTheoryFacilitation2024,lepperUnderminingChildrensIntrinsic1973,CreativityContextUpdate}. Consistent with findings by Serafini et al.~\cite{serafiniEngagingCompanyDevelopers2024}, none of the participants reported discomfort. Instead, they expressed appreciation for the opportunity to reflect critically on the use of LLMs in real-world SOC environments.

\textit{Participants’ Rights} 
Potential participants were first sent an invitation explaining the purpose of the study and emphasizing that participation was entirely voluntary. Those who expressed interest were then asked to review and complete a written consent form (see \Cref{sec:availability}). The consent form described the study process, how data would be collected and used in accordance with the GDPR, how it would be protected, that participants could terminate their involvement in the study at any time without repercussions. After providing written consent, participants completed a short demographic and screening survey. Those who met the study criteria were then invited to take part in an interview. At the start of each interview, participants were reminded of their rights, including the option to withdraw at any time without consequence, to skip questions they felt uncomfortable answering, and to disable the video. Recording began after verbal consent had been confirmed.

\textit{Interactive Task Design (Miro Board)}
To support reflection on LLM integration in SOC workflows, we incorporated interactive tasks using individual Miro~\cite{Miro} boards during the interviews. An alternative approach would have been to conduct these activities purely verbally. However, prior work shows that visual and interactive elicitation techniques support deeper reflection and help participants externalize tacit knowledge~\cite{CreativityContextUpdate,vanbraakElicitingTacitKnowledge2018,roseVisualMethodologiesIntroduction2016}. The tasks were designed to focus on abstract workflows rather than concrete incidents or customer data, and we explicitly instructed participants not to enter any customer-specific, operationally sensitive, or personally identifying information. 

Prior to the interview session, participants received instructions on how to access their individual board, including the option to join via a guest account without creating or logging into a personal Miro account. We also informed them that using a guest account reduces potential privacy risks (i.e., interactions linked to the account), but that Miro may still collect technical metadata, including IP address, browser, and device information. We note that some participants may nonetheless have accessed the board while logged in to their existing Miro accounts. While no directly identifying information (e.g., names or demographic details) was collected within the Miro boards, all participant contributions constitute personal data. Miro processes data in compliance with the \gls{GDPR}~\cite{MiroSecurityCompliance2025}. To further mitigate associated risks, each board was individually password-protected and accessible only to the respective participant and the interviewer. Immediately following each interview, the Miro boards were exported as PNGs and then permanently deleted from the Miro platform to ensure that no potential personal accounts remained linked to the data. Prior to analysis, one researcher reviewed each exported board to confirm that no personally identifiable, customer-specific, or sensitive operational information was present, anonymizing it where necessary. No directly identifying information (e.g., names) was found. The anonymized boards and interview transcripts were stored on secure university servers with restricted access to the research team.

\textit{Data Collection}
Data were collected at multiple stages of the study, including the demographic and screening survey, interview recordings, and artifacts produced during the interactive Miro board tasks. Only anonymized data were retained. We stored all participants' data encrypted using GDPR-compliant services. We used Whisper~\cite{OpenaiWhisper2025} locally for \gls{GDPR}-compliant transcriptions of the recordings. After transcription, one researcher checked all transcripts for errors and ensured all transcripts were anonymized. Once transcription was finalized, we deleted the recording. Naturally, the artifacts related to this study are accessible exclusively to the members directly involved in the study. At no stage do we share any raw or processed data with any third-party entities.

\textit{Participant Benefits}
This study was designed to provide direct and indirect benefits to both participants and the broader SOC community. During the interviews, participants were given structured opportunities to reflect on their own use of LLMs, anticipated risks, and integration strategies. Participants explicitly stated that this process helped them identify previously unconsidered challenges, clarify their assumptions, and articulate organizational concerns related to governance, accountability, and over-reliance. Participants responded positively to the study and expressed appreciation for the opportunity to share their perspectives.

\textit{Benefits to the SOC Community} Beyond the individual benefits, this study aims to provide researchers, tool builders, and SOC decision-makers with aggregated insights into practitioners' expectations, constraints, and risk perceptions regarding LLM adoption. By grounding discussions of LLM integration in practitioners' experiences, our findings will help ensure that future integration is more realistic, human-centred, and operationally safe.

\makeatletter
\begingroup
\setlength{\@fptop}{0pt}
\clearpage
\endgroup
\makeatother

\section{Meta-Review}

The following meta-review was prepared by the program committee for the 2027
IEEE Symposium on Security and Privacy (S\&P) as part of the review process as
detailed in the call for papers.

\subsection{Summary}

This paper presents a qualitative study with SOC practitioners at various levels across multiple companies about their perspectives and experiences with LLMs in the SOC context. The paper aims to understand use cases of LLMs in SOC workflows, the challenges associated with LLM integration, and practitioners’ perceptions of LLM-related risks and countermeasures.

\subsection{Scientific Contributions}
\begin{itemize}
\item Provides a Valuable Step Forward in an Established Field
\end{itemize}

\subsection{Reasons for Acceptance}
\begin{enumerate}
\item This paper addresses a timely and operationally relevant topic: the integration of LLMs into SOC workflows.
\item The inclusion of participants across different SOC roles provides useful perspectives on LLM adoption across organizational levels and responsibilities.
\item The study design is well suited to the research questions, with the Miro boards providing a useful complement to the interviews for eliciting and structuring participants’ perspectives.
\end{enumerate}

\subsection{Noteworthy Concerns} 
\begin{enumerate} 
\item The Miro boards are a strength of the study, but their workflow structure is underused and underexplored in the analysis. Given RQ2’s focus on integrating LLMs into SOC workflows, the paper would benefit from holistically examining how participants viewed the individually-described workflow pieces fitting together, rather than only the siloed components. The paper would further benefit from discussing about how the individual workflow components relate to each other.
\item The paper does not aim to clarify the relationship between LLM integration and SOC maturity. However, given its relevance to understanding when and how LLMs can realistically be integrated into SOC workflows, this is an important missing piece rather than simply an opportunity for future work.
\item The paper attempts to distinguish between experience-based and envisioned claims, but the aggregated representation remains a concern. A finding is marked as experience-based if even one participant had relevant experience, while others may have only envisioned it. This can make a use case or challenge appear more grounded in practice than it actually is.
\end{enumerate}



\end{document}